\PassOptionsToPackage{table,dvipsnames}{xcolor}

\documentclass[sigconf, nonacm]{acmart}

\newcommand\vldbdoi{XX.XX/XXX.XX}
\newcommand\vldbpages{XXX-XXX}

\newcommand\vldbvolume{18}
\newcommand\vldbissue{6}
\newcommand\vldbyear{2025}
\newcommand\vldbauthors{\authors}
\newcommand\vldbtitle{\shorttitle} 
\newcommand\vldbavailabilityurl{URL_TO_YOUR_ARTIFACTS}
\newcommand\vldbpagestyle{empty} 
 \usepackage{cuted}

\usepackage{amsmath,amsfonts}

\usepackage{algorithmic}
\usepackage{textcomp}
 
 \newcommand{\mybox}[1]
{
\vspace{1.8mm}
\noindent \hspace{-1mm} 
\setlength\fboxsep{1mm}
\fbox{\parbox{\dimexpr\linewidth-2\fboxsep-2\fboxrule}{\itshape #1}}
}

\usepackage{booktabs}
\usepackage{hyperref}
\usepackage{caption}
\usepackage[justification=centering]{subcaption}
\usepackage{enumitem}
\usepackage{multirow}
\usepackage{placeins}
\usepackage{outlines}
\usepackage[linesnumbered]{algorithm2e}
\usepackage{amsthm}
\usepackage{xcolor} 
 \definecolor{mygray}{gray}{0.92}
 \definecolor{belyellow}{RGB}{181, 156, 86}

\SetCommentSty{mycommfont}

\newcommand{\revisionone}[1]{\textcolor{black}{#1}}

\newcommand{\revisionthree}[1]{\textcolor{black}{#1}}

\usepackage[title]{appendix}

\usepackage{booktabs}
\usepackage{adjustbox}
\usepackage{blindtext}

\usepackage{ifsym}

\usepackage{braket}
\usepackage{mathtools}

\usepackage[utf8]{inputenc}
\usepackage[english]{babel}

\theoremstyle{definition}

\newtheorem{exmp}{Example}[section]

\newcounter{RihanNOC}
\stepcounter{RihanNOC}

\newcounter{FlorisNOC}
\stepcounter{FlorisNOC}

\newcounter{TimNOC}

\newcounter{ShihNOC}
\stepcounter{ShihNOC}

\newcounter{td}

\newcounter{all}

\newcommand{\para}[1]{\noindent\textbf{#1.}}

\usepackage{tikz}
\usetikzlibrary{quantikz2}
 
\usepackage[framemethod=TikZ,
innerleftmargin=0.1\baselineskip]{mdframed}

\newmdenv[linecolor=gray,backgroundcolor=mygray]{myframe}
\definecolor{darkgreen}{rgb}{0.0, 0.2, 0.13}

\newcommand{\CQSS}{CQSS\xspace}
 
\usepackage{silence}
\usepackage[hang,flushmargin]{footmisc} 

\begin{document}

% \title{Data Management in the Noisy Intermediate-Scale Quantum Era [Vision]}

% \title{Data Management in the Noisy Intermediate-Scale Quantum Era}
% \title{Quantum Data Management in the NISQ Era}
 
\title{Quantum Data Management in the NISQ Era: Extended Version}

%%
%% The "author" command and its associated commands are used to define the authors and their affiliations.

\author{Rihan Hai}
% \orcid{0000-0002-3720-6585}
\affiliation{%
\institution{  Web Information Systems Group \\
Delft University of Technology}
  % \   \institution{Delft University of Technology}
}
\email{r.hai@tudelft.nl}

\author{Shih-Han Hung}
% \orcid{0000-0002-3720-6585}
\affiliation{%
\institution{ Department of Electrical Engineering\\
National Taiwan University 
   %   \institution{National Taiwan University
   }
}
\email{shihhanh@ntu.edu.tw}

\author{Tim Coopmans}
% \orcid{0000-0002-3720-6585}
\affiliation{%
 \institution{ QuTech%Applied Quantum Algorithms \\
 %\revisionone{\slash QuTech
 }
 Delft University of Technology %\revisionone{\slash Delft University of Technology}
   %   \institution{ Leiden University}
}
%\email{t.j.coopmans@liacs.leidenuniv.nl}
\email{t.j.coopmans@tudelft.nl}

\author{\revisionthree{Tim Littau}}
% \orcid{0000-0002-3720-6585}
\affiliation{%
\institution{ Web Information Systems Group \\
\revisionthree{Delft University of Technology}}
  % \   \institution{Delft University of Technology}
}
\email{t.m.littau@tudelft.nl}
% \email{\textcolor{orange}{t.m.littau@tudelft.nl}}

\author{Floris Geerts}
% \orcid{0000-0002-3720-6585}
\affiliation{%
\institution{Adrem Data Lab \\
University of Antwerp}
}
\email{floris.geerts@uantwerp.be}

\begin{abstract}
Quantum computing has emerged as a transformative force in the evolution of  computing technology. Recent efforts have applied quantum techniques to classical database challenges, such as query optimization, data integration, index selection, and transaction management. In this paper, we shift focus to a critical yet underexplored area: \emph{data management for quantum computing}. We are currently in the noisy intermediate-scale quantum (NISQ) era, where qubits, while promising, are fragile and still limited in scale. After differentiating quantum data from classical data, we outline current and future data management paradigms in the NISQ era and beyond. We address the data management challenges arising from the emerging demands of near-term quantum computing. Our goal is to chart a clear course for future quantum-oriented data management research, establishing it as a cornerstone for the advancement of quantum computing in the NISQ era.
\end{abstract}

\maketitle

\pagestyle{\vldbpagestyle}
\begingroup\small\noindent\raggedright\textbf{PVLDB Reference Format:}\\
\vldbauthors. \vldbtitle. PVLDB, \vldbvolume(\vldbissue): \vldbpages, \vldbyear.\\
\href{https://doi.org/\vldbdoi}{doi:\vldbdoi}
\endgroup
\begingroup
\renewcommand\thefootnote{}\footnote{\noindent
This work is licensed under the Creative Commons BY-NC-ND 4.0 International License. Visit \url{https://creativecommons.org/licenses/by-nc-nd/4.0/} to view a copy of this license. For any use beyond those covered by this license, obtain permission by emailing \href{mailto:info@vldb.org}{info@vldb.org}. Copyright is held by the owner/author(s). Publication rights licensed to the VLDB Endowment. \\
\raggedright Proceedings of the VLDB Endowment, Vol. \vldbvolume, No. \vldbissue\ %
ISSN 2150-8097. \\
\href{https://doi.org/\vldbdoi}{doi:\vldbdoi} \\
}\addtocounter{footnote}{-1}\endgroup

\ifdefempty{\vldbavailabilityurl}{}{
\vspace{.3cm}
\begingroup\small\noindent\raggedright\textbf{PVLDB Artifact Availability:}\\
The source code, data, and/or other artifacts have been made available at \url{https://github.com/InfiniData-Lab/Quantum}.
\endgroup
}

\section{Introduction}
Data management is crucial in our increasingly data-driven world, exemplified by the widespread use of database systems and the rapid advancement of big data systems \cite{abadi2016beckman, stonebraker2024goes}.
 In recent years, the field of computer science has been energized by the transformative potential of quantum technologies.
Quantum computing promises computational capacities far beyond what traditional computers can achieve \cite{nielsen2010quantum}. 
However, quantum computing is still in a nascent stage, i.e., the \emph{noisy intermediate-scale quantum (NISQ)} era, characterized by quantum computers that are constrained by noise and limited numbers of qubits \cite{Preskill_2018}. 

With the ongoing development of quantum computing, new data management challenges naturally emerge. 
The fundamental differences between quantum and classical computing call for novel data representations to effectively preserve quantum information  \cite{nielsen2010quantum, vinkhuijzen2024knowledge}. 
Moreover, many quantum computing tasks  are inherently \emph{data- and computation-intensive} due to the need to handle large-scale quantum states \cite{boixo2018characterizing}, multidimensional quantum data structures \cite{biamonte2020lecturesquantumtensornetworks, young2023simulatingquantumcomputationsclassical}, and error correction codes \cite{fowler2012surface, fowler2009high, gottesman1997stabilizer,terhal2015quantum}. 
For example, simulating large-scale quantum computation on classical computers involves processing vast amounts of quantum information, leading to significant scalability and optimization challenges  \cite{boixo2018characterizing, 8869942, xu2023herculeantaskclassicalsimulation, young2023simulatingquantumcomputationsclassical}.
By addressing data management challenges in the NISQ era, the database community has a unique opportunity to significantly enhance the scalability and reliability of quantum technologies, which will advance both research and real-world quantum applications. 

These new data management challenges are not yet clearly defined, 
despite the recent efforts by the database community. Existing work has explored leveraging quantum computers as new hardware to address classical database challenges, such as query optimization \cite{trummer2016multiple, fankhauser2021multiple, fankhauser2023multiple, SchonbergerSIGMOD22, schonberger2023SIGMOD, schonberger2023quantuma, schonberger2023quantumb,  GroppeBiDEDE23, GroppeQMLBiDEDE23}, data integration \cite{ScherzingerVLDB2023demo}, index selection \cite{10.14778/3681954.3682025, gruenwald2023index}, and transaction management \cite{GroppeIDEAS20, groppe2021optimizing, OJCC}. 
However, fundamental questions about data management in the NISQ era remain unanswered. Specifically, how should we define and manage data in the context of near-term and future quantum advancements? 
Given the unique features of quantum computing, such as superposition and entanglement, what are the new considerations to be addressed for effective data management?
What data structures and data management systems will best support the development of quantum technologies, particularly given a limited number of qubits, noise, and other challenges of the NISQ era?

While recent vision papers, tutorials, and surveys   \cite{ccalikyilmaz2023opportunities, 10597772, yuan2023quantum, yuan2024quantum, winker2023quantum} 
have begun discussing the intersection of data management and quantum computing, they primarily focus on how quantum technologies can accelerate classical database operations. 
In contrast, our work delves into an equally important yet underexplored area: \emph{data management for quantum computing}.  
At the moment, the fundamental concepts, research directions, and problem definitions in this area remain obscure within the database community. This paper initiates the exploration of these critical aspects and lays the foundation for further deeper integration of data management and quantum computing. We will not dive into the basics of quantum computing and information, as ample resources are available \cite{young2023simulatingquantumcomputationsclassical, quist2024advancing,  xu2023herculeantaskclassicalsimulation, biamonte2020lecturesquantumtensornetworks}. Instead, we focus on outlining the vision for data management paradigms in the NISQ era and on identifying potential research challenges that are particularly relevant to the database community, i.e., which could have a significant impact on the advancement of today’s quantum technologies. 

\para{Contributions} 
Our contributions are summarized as follows:

\begin{itemize}
    \item \textbf{Fundamentals}: 
    We first explain quantum data and differentiate it from classical data
    (Sec.~\ref{sec:def}).
    
    \item \textbf{Roadmap}: We present our vision for data management research for quantum computing in three paradigms %research directions
    (Sec.~\ref{sec:roadmap}).
    
    \item \textbf{Research problems}: We elaborate on near-term research questions in data management for quantum computing, and report on preliminary experimental results
    (Sec.~\ref{sec:problems}).
    
\end{itemize}

\section{Data in the quantum era}
\label{sec:def}

\emph{Classical data} is the information that is collected, processed, and stored with traditional computing methods. Today, most of the classical data is stored and queried using database systems such as relational databases, document stores, graph databases, and vector databases \cite{dbms}.
We refer to \emph{quantum data} as information collected and processed using \emph{quantum computing devices}, i.e., computing devices that follow the rules of quantum mechanics to their advantage \cite{nielsen2010quantum}. Quantum data is represented by qubits.
Next, we list key differences between quantum and classical data below to help understand the unique features of quantum data.

\underline{\emph{1. Quantum data is probabilistic.}} 
Unlike a classical bit, which is 0 or 1, a quantum bit can be in \emph{superposition}.\ 
Mathematically, a zero state and a one state may be represented by a unit vector in the \emph{standard basis}. That is, the zero state $\ket{0}$ is represented by the vector $\begin{bsmallmatrix}1 \\
  0 \end{bsmallmatrix}$
and the one state $\ket{1}$ is represented by the vector $\begin{bsmallmatrix}0 \\
  1 \end{bsmallmatrix}$. \ 
A single qubit state, denoted $\ket{\psi}$, may be represented by a superposition, i.e., a linear combination  of $\ket{0}$ and $\ket{1}$: $\ket{\psi}=\alpha\ket{0}+\beta\ket{1}$, 
for some pair of complex numbers $\alpha,\beta$ called \emph{amplitudes}, which satisfy $|\alpha|^2+|\beta|^2=1$. \ 
The probabilistic nature arises when a \emph{measurement} is performed. 
When a measurement is performed on the state $\ket{\psi}$, the outcome 0 is obtained with probability $|\alpha|^2$ and the outcome 1 is obtained with probability $|\beta|^2$, and the state permanently changes to the obtained outcome. 
In classical computing, individual bits can be concatenated to form a bit string, e.g., the three-bit string ``010.''
Similarly, we may concatenate qubits into a multi-qubit state. 
In this case, an $n$-qubit state can be represented as a superposition of $\ket{x}$ %=\ket{x_1}\ket{x_2}\ldots\ket{x_n}$ 
for $x\in\{0,1\}^n$, or equivalently a vector of $2^n$ components. 

\underline{\emph{2. Quantum data is fragile.}} 
Quantum computers are anticipated to outperform classical computers in solving certain problems. 
However, with current quantum technology in the NISQ era, quantum resources remain scarce,  and quantum data is prone to noise.  \ 
\emph{Decoherence} \cite{preskill1999lecture}, is the process where quantum states lose their coherence due to environmental interactions, 
% leads to the transition from pure states to mixed states, 
resulting in the gradual loss of quantum information. 
Quantum noise, resulting from unintended couplings with the environment, can significantly degrade the performance of quantum computers \cite{unruh1995maintaining}. Commonly used noise models, such as amplitude damping, phase damping, bit-flip, and phase-flip, mathematically describe how various types of quantum noise lead to decoherence \cite{nielsen2010quantum}. 

\underline{\emph{3. Quantum data can be entangled.}}
Another difference between qubits and bits is \emph{entanglement} 
\cite{einstein1935can}, which means that multiple qubits are correlated such that measuring the state of one qubit immediately affects the others. 
% A well-known entangled state is the \emph{Bell's state}:
% $ 
% \ket{\Psi}=\frac{1}{\sqrt 2}(\ket{00}+\ket{11})=\frac{1}{\sqrt{2}}\left[\begin{smallmatrix}
% 1\\ 
% 0\\ 
% 0\\ 
% 1
% \end{smallmatrix}\right]
% $, 
% where the knowledge of one qubit determines the other. See Appendix A of our online report \cite{techreport} for more background.
A well-known entangled state is the \emph{Bell's state}:
\vspace*{-0.1cm}
\begin{equation*}	\label{exmp:EPR pair}
\ket{\Psi}=\frac{1}{\sqrt 2}(\ket{00}+\ket{11})=\frac{1}{\sqrt{2}}\left[\begin{smallmatrix}
1\\ 
0\\ 
0\\ 
1
\end{smallmatrix}\right],
\vspace{-0.1cm}
\end{equation*}
where knowledge of one qubit determines the other. \revisionone{Please see Appendix A of our online report \cite{techreport} for more background.} 

\section{Our Vision:  Data Management for Quantum Computing}
\label{sec:roadmap}

To facilitate future research on data management for quantum computing, we first sketch the whole landscape in Fig.~\ref{fig:paradigms}. We categorize this landscape into three distinct paradigms based on how quantum and classical data are transformed and utilized, and the type of hardware involved—whether a quantum or classical computer is employed. 
Our goal is to introduce data management researchers to quantum use cases, and by making a distinction based on the nature and role of data within cases. 

\subsubsection*{\textbf{I Classical simulation of quantum computing paradigm}: classical data represents quantum states and operations.}
This paradigm presents great potential for new database challenges. It focuses on using \emph{classical data to represent and simulate quantum data}. 
 For example, Bell's state from Sec.~\ref{sec:def} is represented by a classical vector. 
This paradigm is more accessible as it relies on classical computers, not quantum ones. 

The representative task in this paradigm is \emph{simulation}. Simulation is the process of emulating quantum computation, enabling researchers to model and analyze quantum processes as if they were operating on actual quantum hardware \cite{young2023simulatingquantumcomputationsclassical, xu2023herculeantaskclassicalsimulation}.\footnote{In this work, by simulation we refer to \emph{classical simulation}. Another related term is \emph{quantum simulation} \cite{doi:10.1126/science.1177838,georgescu2014quantum}, which pertains to simulating quantum mechanics on a quantum computer, a subject studied in physics.}
Simulations are of paramount importance in the NISQ era \cite{Preskill_2018}. Consider, for instance, the concept of quantum supremacy \cite{boixo2018characterizing}, which seeks to demonstrate the superior capabilities of quantum computers over classical computers. Given that large-scale quantum computers are not yet available in the NISQ era, simulation is essential for comparing the scalability of quantum computers with classical ones. Additionally, simulations play a crucial role in the development of new quantum algorithms, allowing researchers to design, debug, and validate the correctness of these algorithms before deploying them on expensive quantum devices. 
Moreover, simulations aid in the development of quantum hardware by evaluating error mitigation schemes, predicting algorithm runtimes, and more \cite{zulehner2018advanced, jones2019quest, zhou2020limits, villalonga2019flexible, azuma2021tools}.
Simulation serves as a foundational tool across key areas of quantum computing, including  quantum supremacy, quantum algorithms, 
quantum hardware, error correction, and the exploration of potential quantum applications \cite{xu2023herculeantaskclassicalsimulation}. 

The potential for database research in this paradigm is immense. 
First, imagine innovative database systems specifically designed to manage classical data that supports simulation. 
Second, efficient caching strategies for expensive operations, such as repeated simulations of quantum circuits with varying error parameters, present another important research direction. 
The third research direction in this paradigm involves the representation of quantum states and operations. 
This is useful for quantum design questions such as optimizing the number of quantum gates or gate depth~\cite{kissinger2020reducing,zulehner2018efficient} and compiling textbook quantum circuits to only include the low-level operations that real quantum devices support~\cite{smith2020open}.
As the representation of a quantum state (a vector) or quantum gate (a matrix) generally grows exponentially with the number of qubits, 
developing efficient methods to store and process these data structures is key to making simulations of quantum computation feasible, either in vector form or through more compressed structures like tensor networks~\cite{biamonte2020lecturesquantumtensornetworks, orus2019tensor} or other advanced representations~\cite{bravyi2017improved,wetering2020zx,zulehner2018advanced, jozsa2008matchgates}.
 
\subsubsection*{\textbf{II Joint Quantum-Classical Computing paradigm}: classical preprocessing \& postprocessing for quantum computing}
This paradigm focuses on the situation involved in most quantum-technology applications: classical computers handle the preprocessing and postprocessing of data for quantum devices, such as quantum chips, quantum sensors, and quantum computers. In this paradigm, the focus is on data which is primarily stored and processed as \emph{classical data}. 
For instance, a classical computer sends instructions (classical data) to a quantum chip, which performs computations and returns measurement outcomes—also classical data—back to the classical computer. Thus, the quantum device takes classical input and produces classical output.

Database research can enhance this process by developing efficient systems for managing, storing, and querying the classical data involved in preprocessing, postprocessing, and iterative feedback loops between classical computers and quantum devices, ensuring seamless integration and optimization of data workflows. For instance, the development of quantum error correction—a critical area in quantum computing—can benefit significantly from efficient graph data analytics techniques, as further discussed in Sec.~\ref{ssec:opp}.

We demonstrate the importance of this paradigm through three key categories of quantum applications: \emph{First}, applications where the quantum computation is completed immediately upon returning a result. 
For example, two separated quantum computers can generate a secure, shared key (password) by encoding bits into qubits, then sending and measuring the qubits 
~\cite{bennett1984quantum,pirandola2019advances}.
This is followed by classical postprocessing on the bitstring (i.e. classical data) that the measurement returns~\cite{elkouss2011information}.
The resulting bitstring is a secret key that can later be used for secure communication.
In the \emph{second} category, the quantum computation is stalled temporarily, and the quantum chip still holds quantum data on which the computation will continue later.
An example is the detection of errors during a quantum computation, where at fixed timesteps during the computation check measurements are performed~\cite{gottesman1997stabilizer,terhal2015quantum}, whose outcomes are then decoded to find out which error has most likely occurred~\cite{maurya2024managing,battistel2023real}.
\emph{Third}, efficient preprocessing is critical for quantum algorithms, which  requires \emph{loading} or \emph{encoding} classical data into the quantum domain, typically by mapping classical bits onto quantum bits. This process often requires encoding classical data into a quantum circuit, either as part of the algorithm~\cite{suenderhauf2024blockencoding} or to output a quantum state with amplitudes representing the classical data~\cite{cortese2018loading}, which is a challenging problem.\looseness=-1

\begin{figure}[t]
\centering
 \includegraphics[width=\linewidth]{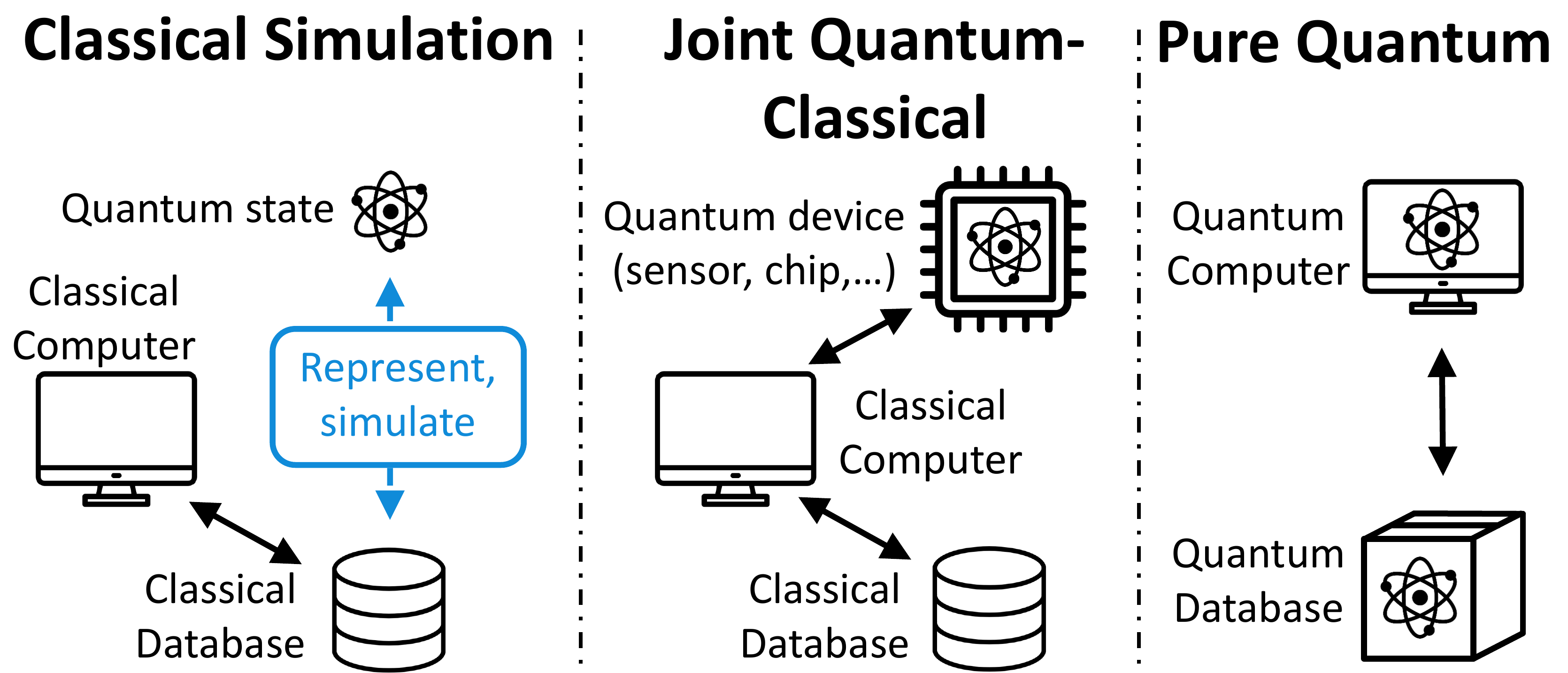}  
\caption{Landscape: Data management for quantum computing}
\label{fig:paradigms}
\Description[]{}
\end{figure}

\subsubsection*{\textbf{III Pure quantum computing paradigm}: quantum-native data storage and processing.}
Finally, we envision a future research paradigm beyond NISQ,  when we have large-scale, fault-tolerant quantum computers.
We will deal with pure \emph{quantum data}, i.e., qubits.
Here, the quantum hardware supports storing the qubits for long enough to perform other tasks in the meantime.
This enables for example a cloud quantum computer~\cite{broadbent2009universal} that rotates its resources among multiple clients, various types of quantum machine learning~\cite{cerezo2022challenges}, and quantum random access memory~\cite{giovannetti2008quantum}, where data can be queried in superposition.
A recent vision \cite{huang2022quantum} shows a possible future:  quantum data is collected from quantum sensing systems, e.g.,  for discovering a black hole; then stored and processed via the quantum memory of a quantum computer.

The data management research in this paradigm will
be centered around handling quantum data, possibly developing quantum-native algorithms. Moreover, storing quantum data will remain challenging, as quantum data storage will still be costly in various ways (the number of qubits is not unbounded, robust storage of quantum data requires large-scale quantum error correction, which is computationally intensive, etc.). 
Another research topic is to efficiently allocate the available qubits.
This includes, for example, various scheduling and allocation tasks~\cite{chandra2022scheduling,fittipaldi2023linear,vanmeter2007system,skrzypczyk2021architecture,gauthier2023control,higgott2021subsystem} and efficient use of different quantum hardware types each of which has speed-decoherence trade-offs~\cite{de2021materials}. Given the amount of quantum data in this paradigm, most research here calls for novel quantum data processing algorithms and systems beyond NISQ era. 
\section{Near-Term Research Problems}
\label{sec:problems}
We will now explore data management research questions across the three paradigms. In Sec.~\ref{ssec:simu2} and \ref{ssec:dbq2}, we focus on a key challenge within Paradigm I: Classical simulation of quantum computing paradigm. In  Sec.~\ref{ssec:opp}, we broaden the scope to include research opportunities across all three paradigms.

\subsection{The Challenge of Simulating Quantum Computation}\label{ssec:simu2}
The input for a simulation is a quantum algorithm described as a quantum circuit,\footnote{A uniform family of circuits, to be precise, one for each input size.} 
consisting of input qubits upon which quantum gates are applied, ultimately resulting in a probability vector of the output states, i.e., measurement outcomes (see Sec.~\ref{sec:def}). Then, a simulation is a classical computation that computes that output distribution (so-called {\em strong} simulation), or samples from it ({\em weak} simulation) \cite{VanDenNes2010}. 
We primarily focus on strong simulation.

\begin{figure}[t]
    \centering

        \centering
        \includegraphics[width=\linewidth]{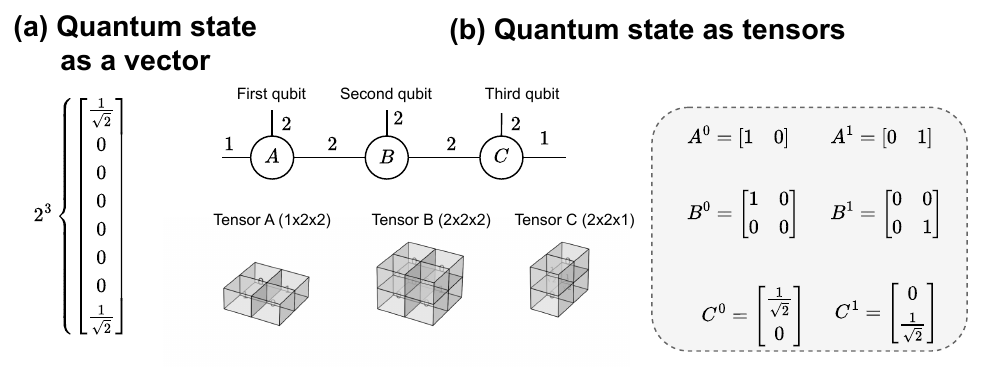}  
        \caption{(a) The 3-qubit  GHZ state  $\frac{1}{\sqrt 2}(\ket{000}+\ket{111})$ represented as a state vector of size $2^3=8$. (b) GHZ state as a tensor network.
        % \cite{perez2007matrix, orus2014practical}. 
        For better visualization, in the grey box, we write the tensors in vector/matrix form by slicing the last dimension of A, B, and C. See Appendices A and B1 of \cite{techreport} for details.}
\label{fig:tensors_example}\Description[]{}
 \end{figure} 

The major practical bottleneck of (strong) simulation is  \emph{scalability} \cite{boixo2018characterizing, 8869942, xu2023herculeantaskclassicalsimulation, young2023simulatingquantumcomputationsclassical}. 
When simulating a quantum state of $n$ qubits, the state is typically represented as a vector with a size of $2^n$. The size of the state vector grows exponentially as $n$  increases. 
For instance,  an experiment demonstrating quantum supremacy required 2.25 petabytes for 48 qubits, reaching the memory limits of today's supercomputers \cite{boixo2018characterizing}.  To overcome the memory restriction, existing simulation tools have explored approximation \cite{vidal2003efficient, medvidovic2021classical, jonsson2018neural}, data compression \cite{wu2019full}, 
parallelization \cite{huang2021efficient},  and distributed computing \cite{smelyanskiy2016qhipsterquantumhighperformance}.  In Fig.~\ref{fig:tensors_example}  we illustrate concepts mentioned in this section.

\subsection{Databases to the Rescue} 
\label{ssec:dbq2}
Simulation offers significant opportunities for database research, and conversely, database expertise can greatly enhance simulation.

\vspace{0.2cm}
\begin{myframe}[roundcorner=3pt,innerleftmargin = 5pt]
We envision a \emph{classical-quantum  simulation system} (\CQSS) with the following capabilities: (i)~automatically providing the most efficient simulation of the input circuit by selecting optimal data structures and operations based on available resources and circuit properties; (ii)~operating inherently out-of-core to support the simulation of large circuits that exceed main memory capacity; (iii)~ensuring consistency to prevent data corruption and enabling recovery in the event of large-scale simulation crashes; and (iv)~improving the entire simulation workflow, including parameter tuning, data collection and querying, exploration and visualization.
\end{myframe}

\medskip
At its core, a \CQSS must, at a minimum,  be capable of evaluating quantum circuits. This primarily involves performing linear algebraic operations,  often described in terms of \emph{tensor networks} \cite{biamonte2020lecturesquantumtensornetworks}.\footnote{For simplicity of exposition, we do not consider non-tensor-based simulation methods which represent quantum states by
their symmetries, rather than by complex-valued vectors \cite{gottesman1998}.} A \emph{tensor} is a multidimensional array, and the dimensions along which a tensor extends are its \emph{indices}.  
A tensor network consists of vertices representing tensors, and edges the indices. 
Free indices are depicted as legs, i.e., edges that only connect to one vertex in the graph. 
Connecting two vertices by joining a leg  corresponds to the contraction with the corresponding indices. 
Fig.~\ref{fig:tensors_example}b shows the tensor network representation of the GHZ state.
Tensor contraction is the summation of shared indices between tensors and is a generalization of matrix multiplication \cite{biamonte2020lecturesquantumtensornetworks}.  See Appendix A of \cite{techreport} for more details.
The first question we ask is as follows.

\begin{figure}[t]
    \centering
    \begin{subfigure}[b]{0.95\linewidth} % Adjusted width to fit three subfigures in one row
        \centering
        \includegraphics[width=0.8\linewidth]{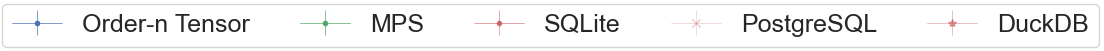}
        \phantomsubcaption
        \label{fig:mem_legend}
     \end{subfigure}
     \setcounter{subfigure}{0}
     \begin{subfigure}[b]{0.48\linewidth} % Adjusted width to fit three subfigures in one row
        \centering
        \includegraphics[width=\linewidth]{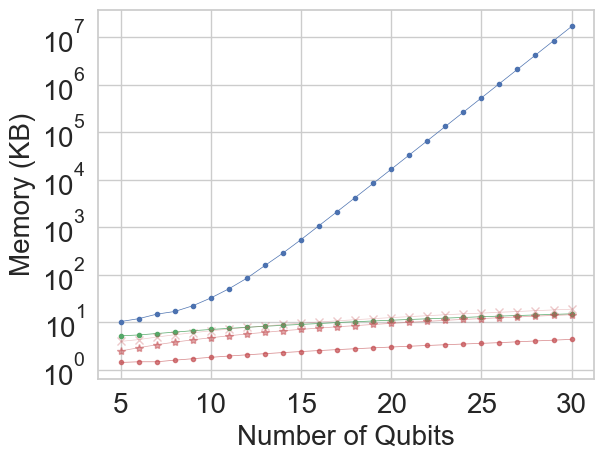}
        \caption{GHZ State (\textbf{sparse} circuit)}
        \label{fig:ghzstate_mem_main}
     \end{subfigure}
    \hfill
      \begin{subfigure}[b]{0.48\linewidth} % Adjusted width to fit three subfigures in one row
        \centering
        \includegraphics[width=\linewidth]{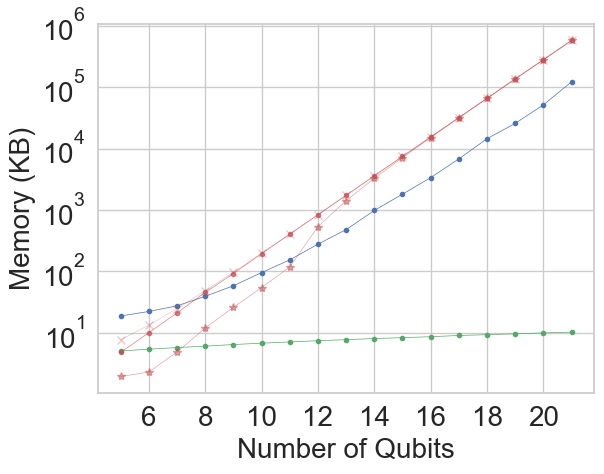}
        \caption{QFT (\textbf{dense} circuit)}
        \label{fig:mem_qft_main}
     \end{subfigure}
          \vspace{-0.2cm}
         \caption{Memory usage comparison of three RDBMS-based solutions (red) against a general approach, i.e., order-n tensor baseline (blue) and the SotA MPS baseline (green), both implemented in Python. The x-axis shows increasing numbers of qubits, and the y-axis shows memory consumption (KB) in $\log_{10}$-scale. For sparse circuit (GHZ), RDBMS solutions demonstrate significantly lower memory usage than the order-n tensor baseline, with SQLite showing a slight advantage over the MPS baseline. For the dense circuit QFT, the memory usage of RDBMS solutions grows comparably to or exceeds the order-n tensor baseline, indicating the need for improvements in handling dense tensor computations.}
    \label{fig:mem_ghz_qft}
     \Description[]{}
\end{figure}

\vspace{0.2cm}
\begin{myframe}[roundcorner=3pt,innerleftmargin = 5pt]
{\bf Q1.} \textit{Should we push the simulation workload to existing DBMSs?}
\end{myframe}

In other words, should we map the simulation workload to SQL queries and store the results in a DBMS? The advantage is that we can benefit from the efficiency and portability of modern database systems. And indeed, initial efforts from the database community \cite{einsteinSQL2023,Trummer24} have begun to address the out-of-core simulation challenge by mapping quantum states and gates to SQL queries. For instance, in \cite{einsteinSQL2023}, qubit states and gates are represented as tensor networks, with sparse tensors in COO format and Einstein summation operations mapped to SQL queries. Additionally, the recent abstract \cite{Trummer24} introduces the idea of supporting dense vectors using SQL UNION ALL and CASE statements. Despite these efforts that support basic quantum states and gates up to 18 qubits and a circuit depth of 18 in separate experiments, the core challenges of representing general quantum states and operations in DBMSs and achieving scalability remain. 
Moreover, our preliminary results in Fig.~\ref{fig:mem_ghz_qft} demonstrate that RDBMS-based simulations are efficient for quantum circuits involving sparse tensor computations. 
All experimental details, %(setting, choice of baselines, results, and analysis), 
together with a space and time complexity analysis, can be found in Appendix B of  \cite{techreport}. An early prototype is described in  \cite{Littau2025Qymera}. 
RDBMS-based solutions are competitive with, and may even outperform, state-of-the-art representations like 
matrix product states (MPS) \cite{perez2007matrix} in specific scenarios. 
Interestingly, this advantage does not extend to dense circuits such as quantum fourier transform (QFT) circuits, highlighting the need for further optimizations to enhance RDBMS systems for simulation workloads. In Appendix C of \cite{techreport} we also discuss the use of NoSQL databases.

\para{Q1 summary} Pushing simulation workloads to DBMSs leverages their efficiency, showing promise for sparse circuits, yet challenges in optimization and scalability remain. 

On the other hand, the connection to tensor computations suggests the following question.

 \vspace{0.1cm}  
\begin{myframe}[roundcorner=3pt,innerleftmargin = 5pt]
{\bf Q2.} \textit{Should we instead leverage tensor-based database technologies for simulation? } 
\end{myframe}
 \vspace{0.1cm}  
Indeed, the database community has a substantial body of work on tensor computation, 
especially with the recent synergies between databases and machine learning (ML) such as \emph{learning ML models over relational data} \cite{Khamis2020, boehm2023optimizing, Makrynioti2019survey, zhou2020database, 10107490}. These studies range from high-level representations to the runtime optimization of ML workflows. At the representation level, numerous studies represent data as tensors \cite{DBLP:journals/pvldb/KoutsoukosNKSAI21} and aim to integrate relational and tensor operations \cite{DBLP:conf/sigmod/SchleichOC16, DBLP:conf/pods/Khamis0NOS18, DBLP:journals/pvldb/HuangSL023}. Meanwhile, system-building efforts from the database community \cite{DBLP:conf/sigmod/GaoLPJ17, DBLP:conf/icde/LuoGGPJ17, DBLP:conf/sigmod/NikolicO18, DBLP:conf/sigmod/SchleichOK0N19, DBLP:conf/cidr/BoehmADGIKLPR20, DBLP:conf/osdi/NakandalaSYKCWI20, DBLP:journals/pvldb/HeNBSSPCCKI22, 10.1145/3514221.3526141, DBLP:conf/cidr/BoehmADGIKLPR20, schleich2023optimizing} focus on performance efficiency and scalability, 
 employing optimization techniques during compilation (e.g., operator fusion \cite{boehm2018optimizing,boehm2023optimizing}) and runtime (e.g., parallelization \cite{DBLP:journals/pvldb/BoehmTRSTBV14}). Similar endeavours exist in the areas of compilers and high performance computing (HPC) \cite{ragan2013halide, kjolstad2017tensor, chen2018tvm, jia2019taso}.

Our envisioned classical-quantum simulation system holds the potential to achieve optimal performance. By integrating relevant optimization techniques from the aforementioned existing systems, we can facilitate the automatic optimization of the underlying tensor network.  The actual runtime performance will depend on several factors: the operations within the tensor network, the sparsity of the tensors involved, the memory layouts (i.e., how tensors are stored in memory, such as row-major or column-major formats), and the available hardware \cite{boehm2023optimizing, schleich2023optimizing}. 
To address challenges unique to quantum computing, we highlight two key gaps below.  

\subsubsection{Quantum-specific optimizations}\label{sssec:opt}
Existing simulators for quantum computing are mostly in Python \cite{cuQuantum, vincent2022jet},  C++ \cite{qsim, vincent2022jet} or Julia \cite{luo2020yao}. There are also dedicated quantum programming languages and domain-specific compilation techniques. See the survey \cite{heim2020quantum}. 
In addition, tensor network rewriting techniques have been considered in the ZX-calculus \cite{Coecke_2011}. 
ZX-calculus is a graphical language for representing and reasoning about quantum computations, represented as tensor networks with specialized tensors called \emph{spiders}  \cite{wetering2020zx}. 
Its graphical nature enables simplifications by locally merging connected vertices while preserving the original tensor, useful in various contexts including formal verification \cite{wetering2020zx,lehmann2022vyzx,de2020zx}. 

Quantum-specific optimizations share some conceptual similarities with classical database optimization, yet they diverge significantly due to the unique characteristics of quantum data and operations. 
First, classical databases rely on relational or NoSQL data models (graph, document-based, etc.) with indexing mechanisms like B-trees or hash tables to facilitate efficient access and retrieval. In simulations, however, quantum states are represented as state vectors (Fig.~2a) or by using more advanced data structures such as tensor networks (Fig.~2b). 
Second, in classical databases, query optimization involves techniques like join ordering to identify the optimal ordering
of join operations between relations for an efficient query plan \cite{steinbrunn1997heuristic, DBLP:journals/pvldb/LeisGMBK015, trummer2017solving}. 
For tensor-based simulators,
a defining feature is the optimization of \emph{contraction order}, 
which determines the most efficient sequence to contract tensors representing quantum states \cite{young2023simulatingquantumcomputationsclassical, cuQuantum}.  
Contraction order optimization parallels the role of join order optimization in traditional databases but is more complex due to the exponential growth of quantum state dimensions and the unique features of quantum data, such as entanglement. 
Third, classical database systems rely heavily on cost estimation to predict and minimize resource use for query execution. Quantum computation simulators, however, lack cost estimation frameworks and instead use optimizations like parallelization, SIMD (Single-Instruction, Multiple-Data) processing, and matrix decomposition techniques (e.g., SVD) to improve performance \cite{young2023simulatingquantumcomputationsclassical}. Quantum-specific issues, such as quantum noise (Sec. 2), complicate simulation further. 
These differences lead to the need for fundamentally new optimization frameworks to manage and simulate quantum computation effectively, beyond classical structures and processes.

\para{Q2 summary} Tensor-based database technologies are promising for boosting simulation performance, with efficiency depending on tensor operations, sparsity, memory layouts, and hardware.

The following question now arises.

\vspace{0.2cm}
\begin{myframe}[roundcorner=3pt,innerleftmargin = 5pt]
{\bf Q3.} \textit{Can we build an optimizer for simulations?} 
\end{myframe}

An optimizer can be designed, given a circuit, to determine the most efficient way to simulate it. For example, when a circuit allows for a tractable simulation (e.g., circuits of bounded treewidth), the optimizer should compile it into an efficient simulation. Similarly, based on the sparsity of (intermediate) quantum states, specialized sparse gate computations may be employed instead of the standard dense computations. 
Particular to the quantum setting is the noisy character of the quantum computation, which is known to impact the efficiency of simulation \cite{Aharonov23,Qi2020,pan2021simulatingsycamorequantumsupremacy,huang2020classical,huang2021efficient}. Hence, an optimizer needs to take noise levels into account.
Moreover, the optimizer should consider the downstream task -- whether it's computing precise probabilities for strong simulations or sampling measurement outcomes for weak simulations -- and select the appropriate simulation strategy (algorithm) accordingly.

As part of the optimizer, we may also want to check whether two quantum states are exactly or approximately equivalent, analogous to checking equivalent conjunctive queries \cite{Chandra+77, Kolaitis+98, cohen2007deciding}. Having algorithms in place for testing equivalence may help to avoid redundant computations during simulation.

\para{Q3 summary} Quantum-specific optimizations require a dedicated simulation optimizer to handle tensor contraction, state sparsity, noise levels, and task-specific strategies, while also eliminating redundant computations through equivalence checking.

% \vspace{-2ex}
\subsubsection{Quantum-specific data representations} \label{subsubsec:data-rep}Simulating quantum circuits  requires precise and efficient representations of quantum states and operations, accounting for the complex nature of quantum mechanics. Ensuring accuracy and scalability as the system grows in complexity is crucial. We mentioned tensor networks as a way to represent the simulation, but various other data representations exist \cite{vinkhuijzen2024knowledge}, e.g., based on algebraic decision diagrams.

\vspace*{0.2cm}
\begin{myframe}[roundcorner=3pt, innerleftmargin = 5pt]
{\bf Q4.} \textit{What are good data representations -- possibly beyond tensors -- for supporting simulations?} 
\end{myframe}

An operation on a data structure is \emph{tractable} if it executes in polynomial time with respect to the input size \cite{darwiche2002knowledge}. 
The choice of data representation can significantly impact whether certain quantum operations (e.g., gates, measurement explained in Sec.~\ref{sec:def} and Appendix A of  \cite{techreport}) are tractable \cite{vinkhuijzen2024knowledge}. 
We emphasize three main requirements for data representation: \emph{expressiveness}, \emph{closure}, and \emph{succinctness/tractability}.

For expressiveness, the data representation must be sufficiently rich to represent quantum states and measurements. 
Closure refers to the property that the result of quantum operations on the represented states can also be represented within the same data representation. This property ensures compositionality.
Finally, it is crucial to minimize the space required to represent a quantum state, as the number of qubits and the level of entanglement can result in exponentially large states. Therefore, we seek data representations that are succinct, minimizing the space needed to store data while still allowing for efficient query processing. It may also be interesting to consider these questions beyond tensor network representations as well
\cite{bravyi2017improved,wetering2020zx,zulehner2018advanced, jozsa2008matchgates}.
A concrete idea for a novel data representation is to combine MPS and the recently-proposed \emph{local-invertible map decision diagram} (LIMDD)~\cite{vinkhuijzen2023limdd}. LIMDD compresses a state vector by lumping together parts of the vector that are equivalent modulo simple quantum gates.
In polynomial time and space in the number of qubits, LIMDDs can simulate circuits that MPS cannot and vice versa~\cite{vinkhuijzen2024knowledge}.
One approach to combine these strengths in a single data structure is applying the lumping to the MPS matrices instead of quantum state vectors, by building upon existing work on equivalence characterizations of MPS~\cite{PhysRevLett.133.010602}.
 
\para{Q4 summary} Effective data representations for quantum simulations require a balance of expressiveness, closure, and succinctness.

\vspace*{0.2cm}
\begin{myframe}[roundcorner=3pt, innerleftmargin = 5pt]
{\bf Q5.} \textit{Are there other benefits from relying on database technology?}
\end{myframe}

By using database technology we aim for the simulation of complex circuits that require significant memory, without relying on HPC infrastructures or the operating system's memory management. By controlling what data is pushed to secondary storage during simulations, we can achieve more efficient I/O behavior. Additionally, transaction management and recovery become important—where a transaction could, for example, represent a sequence of quantum gates—ensuring that computations can restart from saved checkpoints while maintaining the correctness of partially saved results. Parallel evaluation strategies in quantum circuit simulations also highlight the need for effective transaction scheduling, leading to more reliable computational processes. 
Undoubtedly, many challenges remain in this context when considering quantum simulation. We prioritize Q1–Q4, however, as first steps. A more extensive discussion can be found in Appendix D in~\cite{techreport}.

\subsection{More Data Management Opportunities} 
\label{ssec:opp}
Beyond simulation for quantum computing,
quantum-related research is broad, e.g., error correction \cite{shor1995scheme, PhysRevLett.77.793, gottesman1997stabilizer}, %quantum mechanics \cite{}, quantum chemistry \cite{mcardle2020quantum}, 
quantum networks \cite{wehner2018quantum, dahlberg2019link}. Next, we expand our discussion to uncover more opportunities spanning all three paradigms shown in Fig.~\ref{fig:paradigms}.

\para{Quantum error correction \& Graph analytics} 
Quantum error correction (QEC) enables reliable execution of quantum computation on noisy quantum processors and is a crucial part of the roadmap to fault-tolerant quantum computation~\cite{terhal2015quantum}.
QEC process consists of two main aspects: coding and decoding.
The input quantum information as well as the operations to be performed are coded using a quantum error-correcting code such as the surface code \cite{fowler2012surface, bravyi2018correcting, fowler2009high}. Intermittently, decoding is applied: errors are detected and corrected.
Decoders ~\cite{fowler2015minimum, iolius2024decodingalgorithmssurfacecodes, wu2022interpretationunionfinddecoderweighted} often leverage graph theory, solving tasks like a minimum-weight perfect matching (MWPM) problem on a graph where each node represents a check measurement. 
For MWPM, the number of nodes is of the same order as the number of qubits.
Many interesting research problems arise how to model such graphs and design the data formats. 
Moreover, QEC has to be performed at high speeds to avoid decaying quantum states over time (see Sec.~2), e.g. at most several microseconds for some types of quantum hardware for decoding~\cite{battistel2023real}.
The scalability to many qubits, and speed required for QEC might benefit from advanced data management tools and techniques \cite{10.14778/3611540.3611577, yan2005substructure, 10.1145/3448016.3452826, 
10.1145/2627692.2627694,
behnezhad2023exponentially}, offering a promising direction for further exploration.

\para{Quantum experiments \& Scientific data management} Quantum experiments, like any other scientific experiments, generate data and require data management. 
For instance, quantum mechanics experiments on supercomputers often apply HDF5 files to store data \cite{de2011implementing}.
Another compelling direction is to explore how to build specialized data lakes \cite{nargesian2019data, hai2023data} or lakehouses \cite{armbrust2020delta, armbrust2021Lakehouse, jainanalyzing} to support scientific data management for quantum computing.

\para{ML \& Quantum data management} i) \underline{Simulation}: as explained in Sec.~\ref{sssec:opt}, a key challenge in optimizing tensor network based simulation is to find optimal tensor contraction order.  A recent research direction is to apply machine learning (ML), specifically reinforcement learning (RL) and graph neural networks (GNN), to optimize tensor contraction order, addressing the computationally intensive nature of this challenge \cite{pmlr-v162-meirom22a, liu2023classical}. 
ii) \underline{Error correction}: decoding methods for quantum error correction like MWPM, face scalability challenges in large quantum systems, where rapid error detection within strict time limits is essential  \cite{10.1145/3579371.3589037}. An interesting new direction is \emph{data-driven QEC}, which employs ML techniques to   quantum error correction, such as RL \cite{PhysRevLett.131.050601, nautrup2019optimizing, andreasson2019quantum}, multilayer perceptrons (MLP) \cite{chu2022qmlp}, convolutional neural networks \cite{Chamberland_2023}, and GNN \cite{lange2023datadrivendecodingquantumerror}.
iii) \underline{Improving quantum algorithm efficiency:} Quantum algorithms are represented and implemented as quantum circuits, where efficiency can be improved by reducing costly gates like SWAP gate or by minimizing circuit depth and gate count.
ML methods, specifically RL \cite{fan2022optimizing, fösel2021quantumcircuitoptimizationdeep} and MLP \cite{paler2023machine},  have shown potential in optimizing circuit design to enhance the efficiency of quantum circuits on real devices.

\section{Conclusion}
We are at a privileged time in the evolution of data management, closely aligned with the rise of quantum computing. This convergence calls for innovative approaches to data representation, processing, and querying that are compatible with quantum computing. 
We here identify the unique features of quantum data compared to classical data, and advocate the  exploration of three data management paradigms, which reveal a rich field of complex and significant challenges. 
As we continue to explore these paradigms, it becomes clear that the challenges we face, such as enhancing simulations with database technologies, are just the beginning. There exists a broader spectrum of challenges that remain unexplored, which require sustained and focused efforts from both the data management and quantum computing communities. 

\begin{acks}
This publication was supported (in part) by Dutch Research Council (VI.Veni.222.439). 
SHH acknowledges support from the National Science and Technology Council of Taiwan under Grant No. 114-2222-E-002-002-MY3, and from the National Taiwan University Core Consortium Project under Grant No. NTU-CC-114L895503.
TC acknowledges the support received through
the NWO Quantum Technology program (project number NGF.1582.22.035).
\end{acks}

 \bibliographystyle{ACM-Reference-Format}
\bibliography{ref.bib}

% \clearpage
% % \input{letter}

 \clearpage
  
\begin{strip}
\begin{minipage}{2\columnwidth}
\centering\Huge\bf Appendices  
\end{minipage}
\end{strip}

\setlength{\textheight}{\dimexpr\textheight+2\baselineskip\relax}
\begin{figure*}
\centering
    \includegraphics[width=1.0\linewidth]{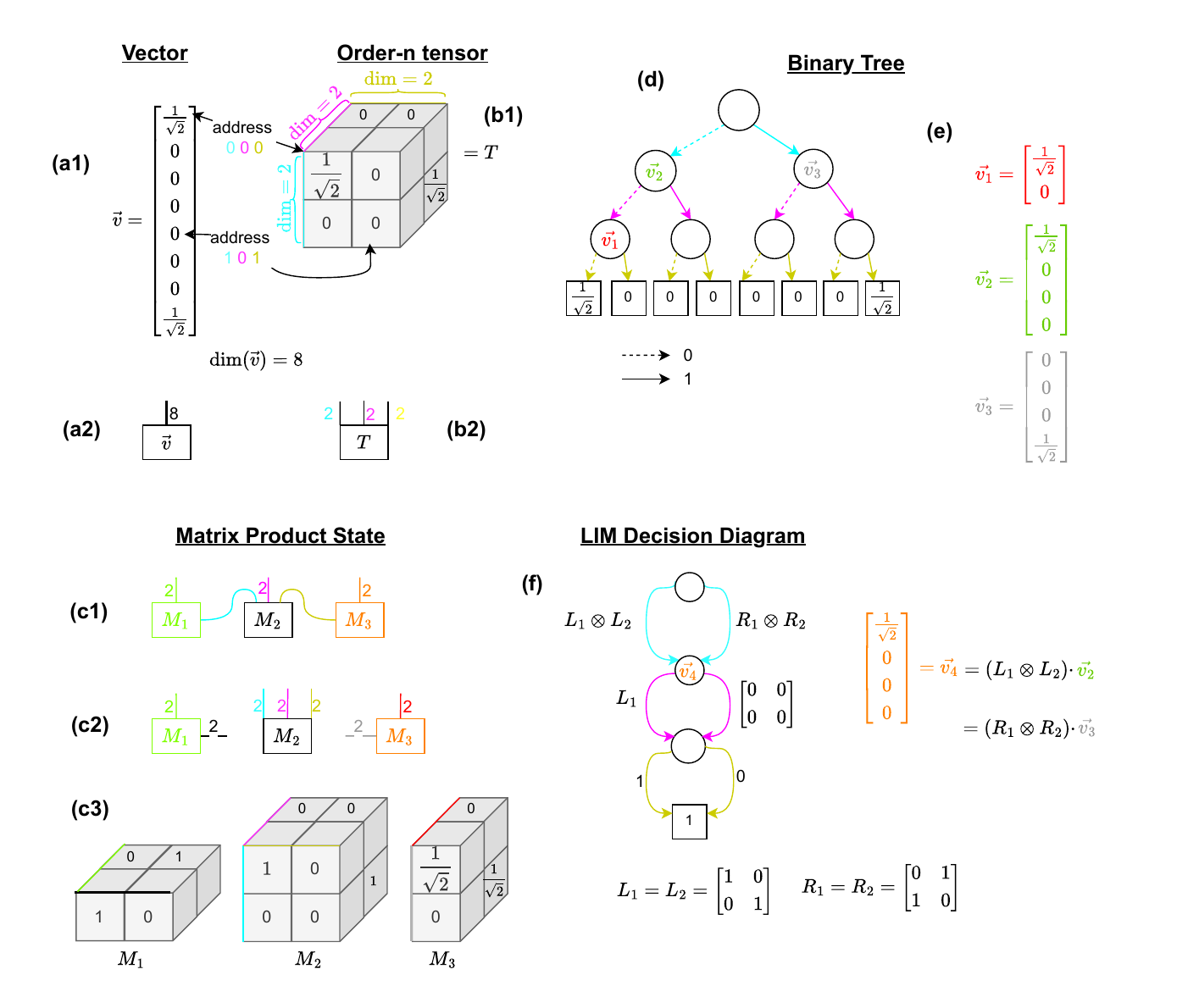}
    \vspace{-1cm}
	\caption{
		Various data structures representing the 3-qubit GHZ state.
		(a1) As vector of $2^3=8$ entries. Each entry is labeled a 3-bit address. The $8$ entries can also be stacked differently as is done in the order-3 tensor $T$ (b1).
		The order-3 tensor $T$ has dimensions $2 \times 2 \times 2$, yielding precisely the $2^3=8$ different addresses.
		(b2) graphically depicts $T$ as a node labeled with the letter $T$, while three outgoing edges (called `legs') indicate the dimensions of the tensor.
		The length-8 vector from (a1) is a single-dimensional tensor, hence its depiction as a node with a single leg in (a2).
%		In the tensor network literature, a tensor is typically depicted as a node labelled with the tensor itself, while three outgoing edges (called `legs') indicate the dimensions of the tensor.
%		E.g. for the order-3 tensor, this is a node with three outgoing edges, each labelled with dimension 2 (b2).
%		Since the length-8 vector can be interpreted as a single-dimensional tensor, its drawing has a single outgoing edge, labelled with dimension 8 (a2).
%		For an $n$-qubit state, the total number of entries in the vector (or equivalently, order-$n$-tensor) equals $2^n$.
%		One potentially more succinct representation is found by decomposing the vector as multiple tensors of often smaller dimensions as is done in the Matrix Product State formalism (c) (coloring consistent with a,b).
		Alternatively to vectors/order-$n$-tensors, (c) the Matrix Product Formalism writes the $n=3$-qubit GHZ states as a collection of $n=3$ tensors $M_1, M_2$   and $M_3$ (c3) %(depicted graphically in c2, b2), 
        some of whose legs are connected (c1).
%		A potentially more succinct representation of the $2^n$ entries as in (a, b) is found by decomposing the vector as multiple tensors of often smaller dimensions as is done in the Matrix Product State formalism (c) (coloring consistent with a,b).
%		The figure depicts three tensors $M_1, T, M_2$, both by specifying their tensors (c3 for $M_1, M_2$ and b1 for $T$), also in graphical notation (c2, b2).
		Converting back to the vector or order-$n$-tensor representation is done by \emph{contracting} the connected legs (a generalization of matrix multiplication; see Sec.~\ref{subsec:MPS}).
		%\textcolor{blue}{Tim: TODO specify where in main text}
%		Tensor contraction is a generalization of matrix multiplication and for the example in the figure, the entry of $\vec{v}$ at address $ijk$ for $i, j, k\in \{0, 1\}$ is found as $\sum_{m=0}^1 \sum_{\ell=0}^1 (M_1)_{im} \cdot T_{mj\ell} \cdot (M_2)_{\ell k}$  (for general definition see text).
%		\textcolor{blue}{Tim: should add coloring to the indices}
%		In the Matrix Product State formalism, the graphical depiction as in (c1) always is a line.
%		In general, more complex graphs are possible, to which we refer as \emph{tensor networks}.
%		An alternative way of compressing the vector/order-$n$-tensor representation is through the use of \emph{decision diagrams}.
		(d) depicts the vector entries from (a) as the leaf nodes of a decision tree where the decisions are indicated by the $n$ address bits (coloring consistent with a,b).
		Each node represents a vector (e), with the root node depicting the original $3$-qubit GHZ-state vector from (a).
		A decision diagram compresses the decision tree by realising that not all of the $2^n$ nodes in the decision tree need to be stored separately.
		The Local-Invertible Map (LIM) Decision Diagram in particular (f) merges nodes since the vectors they represent can be mapped to each other using Kronecker products (denoted $\otimes$) of $2\times 2$ matrices. For LIMDD, the factor $\frac{1}{\sqrt{2}}$ that should be multiplied with each entry is kept track of separately (not displayed).
	\label{fig:datastructures}
	}\vspace{-0.3cm}\Description[]{} 
\end{figure*}

\begin{appendices}
\section{Background: Data representation for Simulation}
\label{sec:defs}
% \floris{references should be added above.}
% \tim{is this comment by Floris' outdated?}
% \rihan{Final check: appendix A includes necessary references} 

We here describe several existing ways to represent a quantum state.
%(Secs.~\ref{sec:state-vector}-\ref{sec:other-representations}).
We also explain how to simulate a quantum circuit which outputs some desired quantum state; the circuit thus consists of a canonical input state (defined below), followed by the application of gates (unitary matrices).
See Fig.~\ref{fig:datastructures} for examples for the main ones mentioned in this work.
%We finish in ???

%The canonical representation is as a \emph{state vector}, and all the other representations could be thought of as compressed versions of the state vector.

\subsection{State vector}
\label{sec:state-vector}
The canonical representation of an $n$-qubit quantum state is the \emph{state vector}: a length-$2^n$ vector $\vec{v}$ of complex entries $v_j$ satisfying the normalization constraint $\sum_{j=0}^{2^n - 1} |v_j|^2 = 1$, where $|.|$ denotes the norm or modulus of a complex number~\cite{nielsen2010quantum}.
See Fig.~\ref{fig:datastructures}(a).
The number $v_j$ is called the amplitude corresponding to the index $j \in \{0, 1, \dots, 2^n - 1\}$.
The index $j$ can be written as length-$n$ bitstring, and we say that the quantum state $\vec{v}$ denotes an amplitude distribution over the length-$n$ bitstrings.
We can also interpret a bitstring as an address, where the $k$-th bit $0$/$1$ recursively indicates chosing the top/bottom half of a length-$2^k$ vector for some $k\in \{0, 1, \dots, 2^n - 1\}$ (see Fig.~\ref{fig:datastructures}(a1) for examples).
Each additional qubit thus adds a single bit to the address, consequently doubling the size of the state vector.

The canonical input state to an $n$-qubit quantum circuit is the length-$2^n$ vector whose topmost entry is $1$, and the remainder entries are $0$.
Simulating a sequence of gates is then done by iteratively updating the quantum-state vector by multiplying it with the next gate (unitary matrix) in line.
The computational complexity of a single gate update is thus the same as multiplying an $N \times N$ matrix with a length-$N$ vector, where $N=2^n$. This is generally $O(N^2)$.
%\tim{can this be improved using the matrix-multiplication exponent $\omega \approx 2.37\dots$?}.

%\tim{add definition, and copy ghz example from main text. Explain again why size grows exponentially with number of qubits}
%A quantum state can be generally expressed in vector form with size grows exponentially in the number of qubits \cite{nielsen2010quantum}. \ 
%The application of a quantum gate evolves the state by multiplying the corresponding unitary operator on the vector, yielding a new vector representing the output. \ 

%\tim{add how to simulate}

\subsection{Order-\texorpdfstring{$n$}{n} tensor}\label{subsec:tensors}
An alternative, equivalent representation of the quantum-state vector is found by reordering the $2^n$ entries in an $n$-dimensional tensor, where every dimension is $2$.
See Fig.~\ref{fig:datastructures}(b) for an example. 

Evaluating a quantum gate on the order-$n$ tensor is done by performing \emph{tensor contraction} between the corresponding unitary matrix (note that every matrix is an order-$2$ tensor and every vector an order-$1$ tensor).
Tensor contraction, a generalization of matrix-matrix multiplication, is most easily explained by graphically visualizing a tensor as a node, with an edge attached for each order labeled with the corresponding dimension, see Fig.~\ref{fig:datastructures}(a2, b2).

Two edges with the same dimension can be connected; contracting this connection is then defined as follows.
Suppose we are given a $3$-tensor $A$ of dimensions $2 \times 3 \times 4$ and a $4$-tensor $B$ of dimensions $3 \times 6 \times 7 \times 8$ and an edge between these which indicates the second index of $A$ and the first index of $B$. Then contracting the edge is an operation which replaces the two nodes by a single one, which represents a $5$-tensor $C$ of dimensions $2 \times 4 \times 6 \times 7 \times 8$, defined as $C_{ijklm} = \sum_{x=1}^3 A_{ixj} \cdot B_{xklm}$.
The naive computational complexity of tensor contraction is $O(X \cdot d)$ where $X$ is the product of the dimensions of the output tensor and $d$ the dimension over which the contraction is performed.
%\tim{but should be able to improve this using Strassen; reference?}
For multiplying an $a \times b$ matrix with a $b \times c$ matrix, this complexity reduces to the well-known $O((a \cdot c) \cdot b)$.

For the order-$n$ tensor, the update after a gate using tensor contraction is equivalent to the matrix-vector gate update for the state vector representation in the sense that one performs exactly the same multiplication and addition of the individual amplitudes.
The gate update also has the same computational complexity.

%Alternatively, the $2^n$ entries of the vector are stacked differently as an $n$-tensor of dimensions $2 \times 2 \times \dots \times 2$.

\subsection{MPS}\label{subsec:MPS}
In the Matrix-Product State (MPS) formalism, the order-$n$ tensor (or, equivalently, the length-$2^n$ vector), is decomposed as product of matrices~\cite{perez2007matrix}.
Formally, an MPS representing an $n$-qubit state is equivalently defined as a collection of $2n$ matrices $M^0_1, M^1_1, M^0_2, \dots, M^0_n, M^1_n$, such that the entry of the quantum state vector at binary address $x_1 x_2 \dots x_n \in \{0, 1\}^n$ is found as the (single entry of the $1\times 1$) matrix product $\prod_{j=1}^n M^{x_j}_j$.
For this product to be a $1\times 1$ matrix, the input dimension of $M^0_1$ and $M^1_1$ needs to be $1$ (i.e. they are lying vectors), and similarly the output dimension of $M^0_n$ and $M^1_n$ is $1$ (i.e. they are standing vectors).

For example, the following MPS represents the 3-qubit GHZ-state (see its state vector in Fig.~\ref{fig:datastructures}(a1)):
\begin{eqnarray*}
    &M^0_1 = \begin{pmatrix}
        1 & 0
    \end{pmatrix}
    &
    M^1_1 = \begin{pmatrix}
        0 & 1
    \end{pmatrix}
    \\
    &M^0_2 = \begin{pmatrix}
        1 & 0\\
        0 & 0
    \end{pmatrix}
    &
    M^1_2 = \begin{pmatrix}
        0 & 0\\
        0 & 1
    \end{pmatrix}
    \\
    &M^0_3 = \begin{pmatrix}
         \frac{1}{\sqrt{2}} \\ 0
    \end{pmatrix}
    &
    M^1_3 = \begin{pmatrix}
        0\\ \frac{1}{\sqrt{2}}
    \end{pmatrix}
\end{eqnarray*}
For example, the GHZ-state's entry $\frac{1}{\sqrt{2}}$ at the top of the vector, i.e. binary index $000$, is found as the matrix product
$
M^0_1 \cdot M^0_2 \cdot M^0_3 = \begin{pmatrix}
        1 & 0
    \end{pmatrix}
    \cdot
    \begin{pmatrix}
        1 & 0\\
        0 & 0
    \end{pmatrix}
    \cdot
    \begin{pmatrix}
         \frac{1}{\sqrt{2}}\\0
    \end{pmatrix}
    =
    \begin{pmatrix}
    \frac{1}{\sqrt{2}} 
    \end{pmatrix}
$
while the entry with value $0$ at binary index $101$ equals
$
M^1_1 \cdot M^0_2 \cdot M^1_3 = \begin{pmatrix}
        0 & 1
    \end{pmatrix}
    \cdot
    \begin{pmatrix}
        1 & 0\\
        0 & 0
    \end{pmatrix}
    \cdot
    \begin{pmatrix}
        0\\ \frac{1}{\sqrt{2}}
    \end{pmatrix}
    =\begin{pmatrix}0\end{pmatrix} 
$

The largest matrix dimension among the matrices $M$ is called \emph{bond dimension}.
In the GHZ example above, the dimensions are $1\times 2$, $2 \times 2$ and $2 \times 1$, so the bond dimension is $2$.
In fact, the bond dimension for the $n$-qubit GHZ state for any positive integer $n$ is $2$, so that the total number of matrix entries that define the MPS is roughly $2 \cdot 2n$, which is much less than the $2^n$ entries that are needed when representing the GHZ state as state vector.
%The study of which states are similarly succinct MPS is a developed one \tim{add reference}.
% ~\cite{}.

Equivalently to the matrix collection above, we can write an $n$-qubit MPS as a collection of $2$ order-$2$ and $n-2$ order-$3$ tensors, whose edge connection structure forms a line where the order-$2$ tensors are at the line's two endpoints.
Contracting all connections then yields the quantum-state vector.
The tensors are obtained by stacking the two MPS matrices for qubit $j \in \{1, 2, \dots, n\}$ such, that they form a single tensor.
For the $3$-qubit GHZ state example above, the $3$ tensors ($M_1$, $M_2$, $M_3$) 
are given in Fig.~\ref{fig:datastructures}(c3), and have dimensions $2 \times 2$, $2 \times 2 \times 2$ and $2 \times 2$.
% We can also interpret them all as order-$3$ tensors by assigning an additional dimension of size $1$ to the order-$2$ tensors.

\begin{figure}[h]
\centering
 \includegraphics[width=0.8\linewidth]{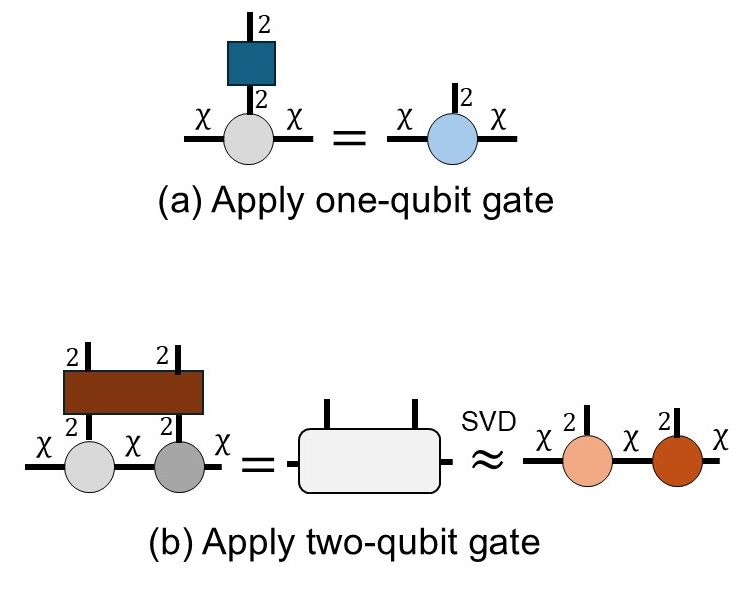}  
 \vspace{-0.3cm}
\caption{Applying gates to qubits in MPS for a fixed (capped) bond dimension $\chi$. (b) A two-qubit gate can increase the bond dimension quadratically; by artificially reducing this back to $\chi$, the output state is approximated.
}
\label{fig:mps_gates}\Description[]{}
\vspace{-0.3cm}
\end{figure}

Simulating~\cite{vidal2003efficient} with MPS starts by constructing the MPS for the canonical input state, which has bond dimension $2$ for any number of qubits $n$.
To update an MPS of bond dimension $\chi$ after a single-qubit gate on qubit $j$, we note that the gate is an order-$2$ tensor (i.e. a matrix) of dimensions $d_{\textnormal{in}}=2$ and $d_{\textnormal{out}} = 2$.
Next, one adds the tensor graphically and connects the $d_{\textnormal{in}}$ edge to the size-$2$ edge of the $j$-th tensor in the MPS, see Fig.~\ref{fig:mps_gates}(a). Then, one contracts the connection, which updates the $j$-th tensor of the MPS only.
The contraction requires accessing all $O(\chi^2)$ entries of the tensor, yielding a total time complexity of $O(\chi^2)$~\cite{vidal2003efficient}.

Applying a two-qubit gate on adjacent qubits starts similarly and requires contracting tensors into an intermediate order-4 tensor, which has time complexity $\mathcal{O}(\chi^2)$.
See Fig.~\ref{fig:mps_gates}(b).
To bring this back to two order-3 tensors, the Singular Value Decomposition (SVD) is applied.
The bond dimension can increase at most quadratically during this process, i.e. from $\chi$ to $\chi^2$.
However, one can choose at this point to make the simulation \emph{approximate} by dropping the terms with small contribution (i.e. small singular value) in order to reduce the bond dimension.
Another approach at approximate simulation is to fix the maximal bond dimension $\chi$ and only keep the largest $\chi$ singular values.
The SVD step dominates the cost of the two-qubit operation, and has a complexity of $\mathcal{O}(\chi^3)$.

%Due to the simple graph structure of MPS, specialized simulation algorithms have been found which are fixed-parameter tractable in the bond dimension.
%MPS can be used for both exact simulation (i.e. where the represented output state vector is precisely 
%or approximate simulation.
%Approximate simulation, i.e. where the resulting output state vector is close to the real output state vector, is achieved by approximating the individual matrices using the singular value decomposition.\tim{add what is gained during approximating matrices: space, smaller bond dimension}
%\tim{should still add citations}
%\rihan{can we here explicitly say MPS only has order-2 and order-3 tensors; also, how does contraction works for MPS? I worry now the time complexity for W/GHZ in MPS/MPO are not very easy to understand}

%A different quantum-state representation

%Unfolding the tensor network definition, an MPS representing an $n$-qubit state is equivalently defined as a collection of $2n$ matrices $M^0_1, M^1_1, M^0_2, \dots, M^0_n, M^1_n$, such that the entry of the quantum state vector at binary address $x_1 x_2 \dots x_n \in \{0, 1\}^n$ is found as the matrix product $\prod_{j=1}^n M^{x_j}_j$.
%(Recall that for matrices, i.e. $2$-tensors, tensor contraction is the same as matrix multiplication)

%\tim{Tim should continue writing/polishing the writing from here on}

\subsection{Other representations}
\label{sec:other-representations}
A \emph{tensor network}~\cite{orus2019tensor} is a generalization of MPS where the graph structure is not a line but a more complex graph, and individual tensors can be of higher order $2$ or $3$.
Due to these relaxations, tensor networks can succinctly represent states which MPS cannot~\cite{herringer2023classificationof}, at the cost of increased hardness of operations~\cite{o2019parameterization}, and have proven powerful when simulating quantum circuits~\cite{huang2021efficient}.

Next, \emph{decision diagrams}, originally developed for reasoning about binary functions, have also been applied to the quantum domain~\cite{thanos2024automated}.
Decision diagrams can be thought of as a compressed version of a binary decision tree of height $n$, where the $2^n$ leaves contain the entries of the quantum-state, see Fig.~\ref{fig:datastructures}(d,e).
The compression is homomorphic, i.e. operations (such as quantum gates) can be applied on the decision diagram without ever having to unfold to the full tree.
Among the most succinct yet tractable types are Local-Invertible-Map Decision Diagrams (LIMDDs), which achieve the compression by noting that nodes in the tree, which correspond to smaller vectors, need not be stored separately if they can be mapped to each other by simple (invertible) matrices~\cite{vinkhuijzen2023limdd}.
See Fig.~\ref{fig:datastructures}(f).
LIMDDs can succinctly represent and simulate a range of relevant states and circuits~\cite{vinkhuijzen2023limdd,vinkhuijzen2024knowledge}.

The \emph{ZX-calculus}~\cite{wetering2020zx} is a graphical language, inspired by tensor networks.
In the ZX-calculus, a quantum circuit containing gates from some finite-sized elementary gate set is represented by a graph with a node for each gate; each node is labeled with the type of gate it represents.
The ZX-calculus also contains a set of rewrite rules, which for example enable one to prove that two circuits are equivalent.
For quantum-circuit simulation specifically, the typical use is to ask for a specific entry in the quantum state vector that the ZX diagram represents.
The entry is found by attaching a ZX diagram for the input state to the diagram for the circuit, followed by a sequence of rewrite rules and a decomposition which expresses the entry as a linear combination of smaller diagrams which are easier to directly evaluate.

%\subsection{Other representations}

Finally, there are representations of quantum states other than the ones mentioned in the main text, such as 
the extended stabilizer~\cite{bravyi2017improved} and matchgate~\cite{mocherla2023extending} formalisms.
% one based on matchgates~\cite{}.
% We specifically mention the extended stabilizer formalism, which is based on the stabilizer formalism that forms the basis for many techniques in quantum computing, such as measurement-based computation~\cite{}.
% The stabilizer formalism enables polynomial-time simulation of Clifford circuits~\cite{}, and has been extended to universal quantum circuits where the asymptotic runtime is fixed-parameter tractable in the number of non-Clifford gates~\cite{}.
% \tim{todo add citations}

%\subsection{W state and GHZ state}\label{subsec:stateswghz}
%In this section, our aim is to include two simple quantum circuits whose intermediate states are sparse: preparation circuits of W state and GHZ state, which are two different yet representative forms of tripartite entanglements \cite{nielsen2010quantum}. 
%The preparations of the W state and the GHZ state have been found in many application, including secure communications 
%%\shih{can you please add 2 sentences here why W and GHZ state worth our investigation, e.g., maybe they are widely used for xxx---see above}

\begin{figure*}[ht!]
\begin{subfigure}[b]{0.4\textwidth}
    \centering
		\textbf{State}
	\end{subfigure}
	\begin{subfigure}[b]{0.3\textwidth}
		\textbf{Circuit}
	\end{subfigure}
	\begin{subfigure}[b]{0.2\textwidth}
    \centering
		\textbf{Used gates}
	\end{subfigure}
     %
     % W state
     %
	\begin{subfigure}[b]{0.4\textwidth}
		\begin{minipage}{\textwidth}
			\begin{eqnarray*}
				&\ket{W_n} =
				\frac{1}{\sqrt{n}}\big(&\ket{100\dots00} + \ket{010\dots00}
				\\
				&&+ \dots + \ket{000\dots01}\big)
			\end{eqnarray*}
		\end{minipage}
	\end{subfigure}
	\begin{subfigure}[b]{0.3\textwidth}
	\scalebox{0.7}{
    \begin{quantikz}
        \lstick{$\ket{0}$} & \gate{X} & \ctrl{1} & \targ{} &  & &\\
        \lstick{$\ket{0}$} &  &\gate{G(\frac{1}{3})} & \ctrl{-1} & \ctrl{1} & \targ{} & \\
        \lstick{$\ket{0}$} &  & & &\gate{G(\frac{1}{2})} & \ctrl{-1} & \\
    \end{quantikz}
		}
%    \caption{3-qubit W state preparation circuit}
    \label{circ:W}
     \end{subfigure}
	\begin{subfigure}[b]{0.2\textwidth}
    % \begin{eqnarray*}
    %      X &=&  \begin{pmatrix} 0 & 1\\ 1 & 0\end{pmatrix}\\
    %  \end{eqnarray*}
    \vspace{0.5cm}
    \[
     X =
\left(
\begin{matrix}
0  & 1 \\
1 & 0 \\
\end{matrix}
\right)
\]
		\[
G(p) =
\left(
\begin{matrix}
\sqrt{p}  & -\sqrt{1-p} \\
\sqrt{1-p} & \sqrt{p} \\
\end{matrix}
\right)
\]

\[
\text{CNOT} =
\left(
\begin{matrix}
1 & 0 & 0 & 0 \\
0 & 1 & 0 & 0 \\
0 & 0 & 0 & 1 \\
0 & 0 & 1 & 0
\end{matrix}
\right)
\]
\vspace{-2cm}
% \text{CNOT} &=& \begin{pmatrix} 1 & 0 & 0 & 0 \\ 0 & 1 & 0 & 0 \\ 0 & 0 & 0 & 1 \\ 0 & 0 & 1 & 0 \end{pmatrix} 
     \end{subfigure}
	%
	% GHZ State
	%
	\begin{subfigure}[b]{0.4\textwidth}
		\begin{minipage}{\textwidth}
		\begin{eqnarray*}
			\ket{GHZ_n} =
			\frac{1}{\sqrt{2}}\left(\ket{\underbrace{000\dots0}_{n\textnormal{ times}}} + \ket{111\dots1}\right)
			= \begin{pmatrix} \frac{1}{\sqrt{2}} \\ 0 \\ 0 \\ \vdots \\ 0 \\ \frac{1}{\sqrt{2}}\end{pmatrix}
		\end{eqnarray*}
		\end{minipage}
	\end{subfigure}
	\begin{subfigure}[b]{0.3\textwidth}
	\scalebox{0.7}{
    \begin{quantikz}
        \lstick{$\ket{0}$} & \gate{H} & \ctrl{1} &  &\\
        \lstick{$\ket{0}$} &  &\targ{} & \ctrl{1} & \\
        \lstick{$\ket{0}$} &  & &\targ{} & \\
    \end{quantikz}
		}
%    \caption{3-qubit GHZ State preparation circuit}
         \label{fig:ghz-circuit}
     \end{subfigure}
	\begin{subfigure}[b]{0.2\textwidth}
		\centering
			% \begin{eqnarray*}
   %          H &=& \frac{1}{\sqrt{2}} \begin{pmatrix} 1 & 1 \\ 1 & -1\end{pmatrix}

            		\[
H = \frac{1}{\sqrt{2}}
\left(
\begin{matrix}
1  & 1 \\
1 & -1 \\
\end{matrix}
\right)
\]
                 % X = \oplus &=& \begin{pmatrix} 0 & 1\\ 1 & 0\end{pmatrix}
                 % X &=&  \begin{pmatrix} 0 & 1\\ 1 & 0\end{pmatrix}\\
					
%CNOT =
%\begin{bmatrix}
%1 & 0 & 0 & 0 \\
%0 & 1 & 0 & 0 \\
%0 & 0 & 0 & 1 \\
%0 & 0 & 1 & 0 \\
%\end{bmatrix}
			% \end{eqnarray*}
     \end{subfigure}
     %
	%
	% QFT
	%
	\begin{subfigure}[b]{0.4\textwidth}
		\begin{minipage}{\textwidth}
		\begin{eqnarray*}
			\ket{QFT_n} &=
            %QFT_n \cdot %\ket{GHZ_n}
            %\\
            %QFT_n: \ket{x} \mapsto \frac{1}{\sqrt{2^n}} \sum_{j=1}^n x_j w^{x \cdot (k - 1)}
            \frac{1}{\sqrt{2^{n+1}}} \big[&
            (1 + \omega^0)\ket{00\dots 00}  \\
            && + (1 + \omega^{-1})\ket{00\dots01} \\
			&&+ \dots \\
            &&+ 
            (1 + \omega^{-(2^n - 1)})\ket{11\dots11}\big]\\
            &\text{ where } &\omega := e^{-2\pi i / 2^n}
            = \cos(2\pi / 2^n) + i \sin(2\pi / 2^n)
		\end{eqnarray*}
		\end{minipage}
	\end{subfigure}
	\begin{subfigure}[b]{0.3\textwidth}
	\scalebox{0.7}{
    \hspace{-1cm}
    \begin{quantikz}
    \lstick[wires=3]{$\ket{GHZ_3}$}
        & \gate{H} & \gate{R_2} & \gate{R_3} &  & & &\\
         &  &\ctrl{-1} &  &\gate{H} & \gate{R_2} &  &\\
         &  & &\ctrl{-2} &  &\ctrl{-1} & \gate{H} & \\
    \end{quantikz}
    }
%    \caption{3-qubit QFT circuit}
    \label{circ:QFT}
     \end{subfigure}
	\begin{subfigure}[b]{0.2\textwidth}
		\[
R_k =
\left(
\begin{matrix}
	1 & 0\\
	0 & e^{2\pi i / 2^k}
\end{matrix}
\right)
\]
	\end{subfigure}
	\caption{
		Three $n$-qubit quantum states, with circuits that prepare them (the figure gives the $n=3$ quantum circuit as example).
		Depicted matrices represent gates. A thick dot with line indicates a controlled-gate: the gate is only performed on the target qubit if the address bit of the control qubit reads 1.
		(Top) W state~\cite{dur2000three}, the uniform superposition of $n$-bit strings with Hamming weight $1$. The W state is sparse enough since the number of non-zero amplitudes in the standard basis is $n$ but there are $2^n$ component.
		Circuit from \cite{cruz2019efficient}, with total gate count $2n-1$, consisting of $n-1$ controlled-$G(p)$ gates, $n-1$ $CNOT$-gates and one additional $X$ gate.
		(Middle) GHZ state~\cite{greenberger1989going}, whose vector is sparse as it only constains two nonzero entries. Its preparation requires entangling all qubits with each other. We use a circuit with $n$ gates: a single Hadamard and $n-1$ controlled-NOT (CNOT) gates \cite{cruz2019efficient}.
		(Bottom) The output state, denoted here as $\ket{QFT_n}$, of the $n$-qubit Quantum Fourier Transform~\cite{nielsen2010quantum} (QFT) gate, with the $n$-qubit GHZ state (here we use the 3-qubit GHZ state $\ket{\text{GHZ}_3}$) as input.
        \label{fig:ghz-w-qft}
	}    \Description[]{}
\end{figure*}

\section{Preliminary findings}
In this section, we first explain in Sec.\ref{sec:rel} how we represent quantum states in the relational model. Then,
in Sec.~\ref{sec:ana}, we analyze the space and time complexities of the relational representation, comparing it with some of the existing quantum data representations used in our experiments. In Sec.~\ref{ssec:setting} and \ref{ssec:results}, we further detail the experimental setup and report preliminary experimental results.

Both our analytical and experimental evaluations are illustrated using circuits for producing three quantum states: the W state and GHZ states, which found numerous applications in secure communication~\cite{shi2002teleportation, lipinska2018anonymous, miguel2020delocalized, miguel2023quantum} and quantum metrology~\cite{ng2014quantum}, as well as a the output state of a circuit only containing the Quantum Fourier Transform (QFT) gate, a central operation in the famous quantum algorithm by Shor for factoring large integers.
Fig.~\ref{fig:ghz-w-qft} shows the states as well as a circuit for producing them.

\subsection{Relational representation for quantum states}
\label{sec:rel}
We already have seen different representations of quantum states in Section~\ref{sec:defs}.
Here, we describe a simple relational representation based on tensors (Section~\ref{subsec:tensors}). 
In particular, we propose a \emph{sparse} relational encoding of tensors, only storing entries with non-zero values.\footnote{Similar to \cite{einsteinSQL2023}, our relational representation is based on encoding general tensor expressions in a relation schema comprising tensor indices and values. In addition, we extend this representation to incorporate quantum-specific semantics. i.e., treating an n-qubit state as an order-n tensor, and storing only non-zero probability amplitudes. With simple transformations (merging multiple column values into a single text field), our representation can be converted into the qubit state representation used in \cite{Trummer24}.} More precisely, to efficiently represent states using RDBMS, given an $n$-qubit state \( \ket{\psi} \), we encode the non-zero amplitudes and their corresponding basis state indices in a relation $R$ with arity $n+2$, as follows:
\begin{equation*}
\label{eq:relational_matrix}
         R(s_1, s_2, \cdots, s_n, r, c), \text{ for } s_k \in \{0, 1\}, k \in [1, n],  r \in \mathbb{R}, c \in \mathbb{R} ,
\end{equation*}
where:
\begin{itemize}
    \item \( n \) is the number of qubits.
    \item \( s_1, s_2, \cdots, s_n \) represent the binary indices of the basis states in the computational basis (\( \lvert 0 \rangle \) or \( \lvert 1 \rangle \) for each qubit).
    \item \( r \) and \( c \) are the real and imaginary parts, respectively, of the (possibly complex) amplitude associated with the corresponding \emph{non-zero} basis state. Both \( r \) and \( c \) are real-valued.
\end{itemize}

 Hence, a  tuple $t \in R$ is a tuple representing one non-zero basis state and its amplitude. We remark that the total number of tuples in this 
  representation of $\ket{\psi}$ is $\text{nnz}(\ket{\psi})$,  where $\text{nnz}(\ket{\psi})$ 
   denotes the number of non-zero amplitudes in the quantum state $\ket{\psi}$.

% \shih{The relational representation of quantum states here seems different from the graph structure of MPS. In an MPS, the state is represented as a graph, whose nodes represent a subsystem, and whose edges have something to do with the entanglement between subsystems. In a relational representation of databases, vertices are tables and edges are common attributes. I wonder if the correspondence implies that we can benefit from database research.}

\begin{exmp}
In Fig.~\ref{fig:quantum_states}, we illustrate the relational representation of the W, GHZ and QFT states. These states, their circuits, and the matrix representations of the gates used are illustrated in Fig.~\ref{fig:ghz-w-qft}. 
\begin{enumerate}[left=0pt]
\item The 3-qubit W state $\ket{W_3}= \frac{1}{\sqrt{3}} \left( \ket{001} + \ket{010} + \ket{100} \right)$ is a superposition of three basis states with non-zero amplitudes. 
It is represented by three tuples shown in Fig.~\ref{fig:w_state}. For instance, the tuple \( (0, 0, 1, \frac{1}{\sqrt{3}}) \) corresponds to the basis state \( \ket{001} \) with its amplitude \( \frac{1}{\sqrt{3}} \). 
\item The $3$-qubit GHZ state $\ket{\text{GHZ}_3}= \frac{1}{\sqrt{2}} \left( \ket{000} + \ket{111} \right)$ is a maximally entangled state involving only two basis states with non-zero amplitudes. It is represented by two tuples in Fig.~\ref{fig:ghz_state}. 
% % Since the amplitudes in the W and GHZ states are real values, the imaginary column $c$ has only zero values and is therefore omitted in the relational representation.
% \begin{enumerate}[left=0pt,resume]
\item  The output of the 3-qubit QFT circuit, shown in Fig.~\ref{fig:ghz-w-qft}, when given the GHZ state $\frac{1}{\sqrt{2}}(\ket{000} + \ket{111})$ as input, 
is $\ket{\psi_3}=\frac{1}{4} \bigg( 2\ket{000}  +   (1+i)\ket{010} + (1-i)\ket{011} + (1+\frac{1}{\sqrt{2}} + \frac{i}{\sqrt{2}})\ket{100}+ (1-\frac{1}{\sqrt{2}} - \frac{i}{\sqrt{2}})\ket{101} +(1-\frac{1}{\sqrt{2}} + \frac{i}{\sqrt{2}})\ket{110} + (1+\frac{1}{\sqrt{2}} - \frac{i}{\sqrt{2}})\ket{111} \bigg)$. 
We depict its relational representation in  Fig.~\ref{fig:qft_state} consisting of eight tuples.
% The QFT final state involves all eight basis states with only real amplitudes 
% given the GHZ state % $\frac{1}{\sqrt{2}}(\ket{000} + \ket{111}^)$
% as input, which results in the relational 
% shown in 
% Fig.~\ref{fig:qft_state}.
\end{enumerate}
\end{exmp}

\begin{figure}[thb]
    \centering
    \begin{minipage}[T]{0.2\textwidth} % Left side for (a) and (b)
        \begin{subfigure}[t]{\textwidth}
            \centering
            \begin{tabular}{|c|c|c|c|c|}
                \hline
               \(s_1\) & \(s_2\) & \(s_3\) & r & c \\ \hline
                0 & 0 & 1 & \( \frac{1}{\sqrt{3}} \) & 0\\ \hline
                0 & 1 & 0 & \( \frac{1}{\sqrt{3}} \) & 0\\ \hline
                1 & 0 & 0 & \( \frac{1}{\sqrt{3}} \) & 0\\ \hline
            \end{tabular}
            \caption{3-qubit W state }
            \label{fig:w_state}
            \vspace{0.4cm}
        \end{subfigure}
        \vspace{0.5cm} % Spacing between (a) and (b)
        \begin{subfigure}[t]{\textwidth}
            \centering
            \begin{tabular}{|c|c|c|c|c|}
                \hline
               \(s_1\) & \(s_2\) & \(s_3\) & r & 0\\ \hline
                0 & 0 & 0 & \( \frac{1}{\sqrt{2}} \)& 0 \\ \hline
                1 & 1 & 1 & \( \frac{1}{\sqrt{2}} \) & 0\\ \hline
            \end{tabular}
            \caption{3-qubit GHZ state}
            \label{fig:ghz_state}
        \end{subfigure}
    \end{minipage}
    \hfill
    \begin{minipage}[T]{0.25\textwidth} % Right side for (c)
        \begin{subfigure}[t]{\textwidth}
            \centering
            \begin{tabular}{|c|c|c|c|c|}
                \hline
                \(s_1\) & \(s_2\) & \(s_3\) & r & c\\ \hline
                0 & 0 & 0 & \( \frac{1}{2} \) & 0 \\ \hline
                 0 & 1 & 0 & \( \frac{1}{4} \) & \( \frac{1}{4} \) \\ \hline
                0 & 1 & 1 & \( \frac{1}{4} \) & \( -\frac{1}{4} \) \\ \hline  
                1 & 0 & 0 & \( \frac{1}{4}+\frac{1}{4\sqrt{2}} \) & \( \frac{1}{4\sqrt{2}} \) \\ \hline
                1 & 0 & 1 & \( \frac{1}{4}-\frac{1}{4\sqrt{2}} \) & \( -\frac{1}{4\sqrt{2}} \) \\ \hline
                1 & 1 & 0 & \( \frac{1}{4}-\frac{1}{4\sqrt{2}} \) & \( \frac{1}{4\sqrt{2}} \) \\ \hline
                1 & 1 & 1 & \( \frac{1}{4}+\frac{1}{4\sqrt{2}} \) & -\( \frac{1}{4\sqrt{2}} \) \\ \hline
            \end{tabular}
            \caption{Final state $\ket{\psi_3}$ of 3-qubit QFT circuit for GHZ state}
            % \centering
            % \begin{tabular}{|c|c|c|c|}
            %     \hline
            %     \(s_1\) & \(s_2\) & \(s_3\) & r \\ \hline
            %     0 & 0 & 0 & \( \frac{1}{\sqrt{8}} \) \\ \hline
            %     1 & 0 & 0 & \( \frac{1}{\sqrt{8}} \) \\ \hline
            %     0 & 1 & 0 & \( \frac{1}{\sqrt{8}} \) \\ \hline
            %     1 & 1 & 0 & \( \frac{1}{\sqrt{8}} \) \\ \hline
            %     0 & 0 & 1 & \( \frac{1}{\sqrt{8}} \) \\ \hline
            %     1 & 0 & 1 & \( \frac{1}{\sqrt{8}} \) \\ \hline
            %     0 & 1 & 1 & \( \frac{1}{\sqrt{8}} \) \\ \hline
            %     1 & 1 & 1 & \( \frac{1}{\sqrt{8}} \) \\ \hline
            % \end{tabular}
            % \caption{Final state $\ket{\psi_3}$ of 3-qubit QFT circuit for initial state $\ket{000}$}
            \label{fig:qft_state}
        \end{subfigure}
    \end{minipage}
    \vspace{-2ex }
    \caption{Relational representations of different quantum states.
    % : (a) W state, (b) GHZ state, and (c) QFT state.
    }
    \label{fig:quantum_states}\Description[]{}
\end{figure}

\begin{table*}[ht]
\centering
\begin{tabular}{|c|cc|c|}
\hline
% \rowcolor{gray!20} \textbf{Type}                             & \multicolumn{2}{c|}{\textbf{Exact representation}}                                             & \textbf{Approximate representation} \\ \hline
\rowcolor{gray!20} \textbf{State}  & \multicolumn{1}{c|}{\textbf{Order-$n$ tensor}}               & \textbf{Relational representation}                         & \textbf{MPS}                                 \\            
\rowcolor{gray!20}       $\ket{\psi}$           & \multicolumn{1}{c|}{\textbf{(Baseline I)}} & \textbf{(RDBMS solutions)}          & \textbf{(Baseline II)}       \\ \hline
\textbf{General state}                                             & \multicolumn{1}{c|}{$\mathcal{O}(2^n)$}             & $\mathcal{O}(n \cdot \text{nnz}(\ket{\psi}))$ & $\mathcal{O}(n \chi^2)$             \\ \hline
\textbf{$\text{W}_n$ State}                                             & \multicolumn{1}{c|}{$\mathcal{O}(2^n)$}             & $\mathcal{O}(n^2)$                       & $\mathcal{O}(n)$                    \\ \hline
\textbf{$\text{GHZ}_n$ State}                                           & \multicolumn{1}{c|}{$\mathcal{O}(2^n)$}             & $\mathcal{O}(n)$                         & $\mathcal{O}(n)$                    \\ \hline
\textbf{$\text{QFT}_n$}                                                 & \multicolumn{1}{c|}{$\mathcal{O}(2^n)$}             & $\mathcal{O}(n \cdot 2^n)$                     & $\mathcal{O}(n 
\chi^2)$             \\ \hline
\end{tabular}
\caption{Space complexity comparison of different representations of state $\ket{\psi}$. Here, $n$ is the number of qubits, $nnz(\ket{\psi})$ denotes the number of non-zero probability amplitudes in the state \( \ket{\psi} \), and the MPS bond dimension $\chi$ is a fixed constant that one chooses oneself, potentially making the representation approximate.}
\label{tab:representation}
\end{table*}

% \medskip
% \noindent
% \emph{Remarks.}  Our
In summary, our relational representation is designed to compactly encode sparse quantum states, significantly reducing storage requirements compared to the trivial representation, which requires $2^n$ tuples. 
% As discussed in Section~\ref{sec:defs}, several alternative compact representations are available. A key challenge lies in integrating these approaches with relational encodings to achieve unified and efficient representations.

% In sparse quantum states, such as the W state and GHZ state, the number of basis states with non-zero coefficients is significantly smaller than the total number of basis states (\( 2^n \)). 
% In these cases, the relational model offers a compact and efficient representation.\footnote{Further optimizations to enhance this representation remain a promising avenue for future research.} 
% However, as illustrated in Fig.~\ref{fig:qft_state}, the benefit of the relational model diminishes for dense states like the QFT final state, where a majority of the basis states have non-zero coefficients. 

% In the following sections, we formally analyze the space and time complexities of the relational representation to further evaluate its applicability.

% The table \( T \) has \( \text{nnz}(\ket{\psi}) \) rows, corresponding to the number of non-zero coefficients in the quantum state. This structure leverages the sparse representation for efficiency, where only the significant components of the state are stored in the relational database, making it suitable for quantum simulation tasks. 

\subsection{Complexity Analysis}
\label{sec:ana}

We compare the space and time complexity of quantum computations across three representations used in our experiments. This section first addresses space complexity and then provides an overview of time complexity. The representations considered are:

\begin{itemize}[left=0pt]
    \item \textbf{NumPy Tensor Format:} A flattened row-major tensor representation. This approach avoids explicitly storing tensor indices, as elements are accessed using stride values along each dimension. The space complexity is measured in terms of the total number of entries.
    
    \item \textbf{Relational Tensor Format:} As just defined, this representation stores indices alongside non-zero values. The space complexity is measured in terms of number of non-zero values times the space per entries ($n+2$).
    
    \item \textbf{Matrix Product State (MPS):} A decomposition of tensors with a bond dimension \(\chi\), where each component is stored in NumPy tensor format (See also Section~\ref{subsec:MPS}).
    The space complexity is the sum of the space complexity of these components.
    For a given state \(\ket{\psi}\), we denote the smallest bond dimension, minimized over all MPS that represent  \(\ket{\psi}\), as \(\chi_{\text{exact}}(\ket{\psi})\).
    We fix \(\chi\) to a constant throughout the simulation, which makes the simulation approximate if for some intermediate quantum state $\ket{\psi}$ in the circuit computation it holds that \(\chi < \chi_{\text{exact}}(\ket{\psi})\).
\end{itemize}

\medskip
\noindent
\textbf{Space Complexity.}
As usual, we evaluate space complexity in terms of the size of the largest tensor encountered during computation. Let \(n\) represent the number of input qubits. A summary of the results is presented in Table~\ref{tab:representation}. It is known that general quantum computations can be carried out using local gates, i.e., gates only affecting a constant number of qubits. We assume local gates throughout this section.

\medskip
\noindent
$\rhd$ \emph{General Case.} For arbitrary quantum computations,space consumption depends on the representation of an arbitrary quantum state \(\ket{\psi}\):
\begin{itemize}[left=0pt]
    \item For NumPy tensors, space complexity is \(\mathcal{O}(2^n)\), as it stores all \(2^n\) elements of the tensor. Indeed, the application of local gates can be done in  \(\mathcal{O}(2^n)\) space in as this corresponds to multiplication with a matrix of fixed (constant) dimensions.
    \item For the relational representation, only non-zero values are stored along with their indices, resulting in \(\mathcal{O}(n \cdot \text{nnz}(\ket{\psi}))\), where \(\text{nnz}(\ket{\psi})\) denotes the number of non-zero elements. Also here, locality of gates ensures that the constant matrix multiplication is carried by a join requiring  \(\mathcal{O}(2^n)\) space.
    \item For MPS, with bond dimension \(\chi\), the representation includes two matrices of size \(2 \times \chi\) and \(n-2\) tensors of size \(\chi \times 2 \times \chi\). Stored in NumPy format, the space complexity is \(\mathcal{O}(n \cdot \chi^2)\).
    %Note that \(\chi\) depends on \(n\) and can grow as large as \(2^n\).
    %Any tensor can be represented in MPS by allowing $\chi=2^{n/2}$.
\end{itemize}

\medskip
\noindent
$\rhd$ \emph{Examples: W, GHZ, and QFT States.}
For all three examples one can verify that the space complexity is dominated by the space complexity of their final quantum state. 
% These are, for the \(n\)-qubit W state:
% \[
% \lvert W_n \rangle = \frac{1}{\sqrt{n}} \left( \lvert 10\cdots 0 \rangle + \lvert 01\cdots 0 \rangle + \cdots + \lvert 00\cdots 1 \rangle \right),
% \]
% and for the \(n\)-qubit GHZ state:
% \[
% \lvert \text{GHZ}_n \rangle = \frac{1}{\sqrt{2}} \left( \lvert 0\cdots 0 \rangle + \lvert 1\cdots 1 \rangle \right).
% \]
The  number of non-zero elements for these states are \(\text{nnz}(\lvert W_n \rangle) = n\) and \(\text{nnz}(\lvert \text{GHZ}_n \rangle) = 2\) and \(\text{nnz}(\lvert \text{QFT}_n \rangle) = 2^n\). Moreover, the $W$ and $GHZ$ states can be represented exactly with bond dimension \(\chi_{\text{exact}}=2\)~\cite{perez2007matrix, vinkhuijzen2024knowledge}. As a consequence, for these three states:
\begin{itemize}[left=0pt]
    \item For NumPy tensors, the space complexity remains \(\mathcal{O}(2^n)\) as sparsity is not leveraged.
\item For the relational representation, the general space complexity is $O(n \cdot \text{nnz}(\ket{\psi})$, so becomes \(\mathcal{O}(n^2)\) for \(\lvert W_n \rangle\) and \(\mathcal{O}(n)\) for \(\lvert \text{GHZ}_n \rangle\), by leveraging sparsity but at the cost of storing indices. For the \(n\)-qubit QFT state, it is $\mathcal{O}(n \cdot 2^n)$.
\item For MPS, as we fix the bond dimension $\chi$, the tensors are all of size $O(\chi^2)$, yielding a combined space complexity of all $n$ tensors of $\mathcal{O}(n \chi^2)$.
However, the $W$ state and $GHZ$ states both have constant bond dimension (\(\chi_{\text{exact}} = 2\)), independently of the number of qubits $n$, so the space complexity is \(\mathcal{O}(n)\) for these two states.
%Pinning down the bond dimension of the \(n\)-qubit QFT state is more complex (in general, the QFT has subtle entanglement generation power~\cite{chen2022quantum}) and we consider it out of scope for this preliminary analysis. Here we represents the upper bounded of QFT as $\chi(|\text{QFT}_n)$ and the space complexity is  $\mathcal{O}(n (
%\chi(|\text{QFT}_n\rangle)^2)$.
% \tim{changed. Actually, I expect it has bond dimension 4...}
\end{itemize}

\medskip
\noindent
$\rhd$ \emph{Remarks.}
This analysis demonstrates the advantages of relational representations over standard numerical formats like NumPy, particularly for sparse quantum states. These advantages are evident in sparse quantum circuits, where the sparsity of the quantum state is preserved, as in state preparation tasks \cite{sparseStates}. 
Currently, the relational representation does not utilize MPS factorization. However, each MPS component could naturally be represented relationally. Similarly, there are connections to treewidth of quantum circuits that could be integrated \cite{Markov+2008}.
Developing more advanced compact representations, leveraging recent work on such representations (see Section~\ref{sec:defs}), represents a key research challenge.

\begin{table*}[ht]
\centering
\begin{tabular}{|c|cc|c|}
\hline
% \rowcolor{gray!20} \textbf{Type}                             & \multicolumn{2}{c|}{\textbf{Exact representation}}                                             & \textbf{Approximate representation} \\ \hline
\rowcolor{gray!20} 
\textbf{State}  & \multicolumn{1}{c|}{\textbf{Order-$n$ tensor}}               & \textbf{Relational representation}                         & \textbf{MPS}                                 \\    
\rowcolor{gray!20}                              $\ket{\psi}$       & \multicolumn{1}{c|}{\textbf{(Baseline I)}} & \textbf{(RDBMS solutions)}          & \textbf{(Baseline II)}       \\ \hline
\textbf{W State}                                             & \multicolumn{1}{c|}{$\mathcal{O}(n \cdot 2^n)$}             &  $\mathcal{O}(n^2)$                      & $\mathcal{O}(n)$                    \\ \hline
\textbf{GHZ State}                                           & \multicolumn{1}{c|}{$\mathcal{O}(n \cdot 2^n)$}             &  $\mathcal{O}(n)$  & $\mathcal{O}(n)$                    \\ \hline
\textbf{QFT}                                                 & \multicolumn{1}{c|}{$\mathcal{O}(n^2 \cdot 2^n)$}             &  $\mathcal{O}(n^3 \cdot 2^n)$                     & $\mathcal{O}(n^3 \chi^3)$             \\ \hline
\end{tabular}
\caption{Time complexity comparison for the circuits from Fig.~\ref{fig:ghz-w-qft} for producing state $\ket{\psi}$. Here, $n$ is the number of qubits, 
%$nnz(\ket{\psi})$ denotes the number of non-zero probability amplitudes in the state \( \ket{\psi} \), 
and the MPS bond dimension $\chi$ is a fixed constant that one chooses oneself, potentially making the representation approximate.
}
\label{tab:time}
\end{table*}

\medskip
\noindent
\textbf{Time complexity.}\label{ssec:time_analysis} 
The time complexity of simulating general quantum circuits is a complex problem and depends on many factors, including the number of gates, circuit depth, type of gates, circuit length, and
the graph structure of the circuit. 
% other factors depending on the specific simulation application. In practice, the actual implementation and hardware setting (e.g., GPU cache size) \cite{vallero2024statepracticeevaluatinggpu}
% can also significantly impact time efficiency. 
For our purposes, and in particular for our experiments, it suffices to provide a comparison of the time complexity of our three example circuits: The W state and GHZ state preparation circuits, and QFT circuits from Fig.~\ref{fig:ghz-w-qft}.
Our analysis is summarized in Table~\ref{tab:time}. 
%
% As a starting point, we focus on the three circuits: W state and GHZ state preparation courts, and QFT circuits. 
% Consistent with the preliminary results in Sec.~\ref{ssec:setting}, we primarily analyze the impact of the number of qubits, gates, and gate types. For simplicity, we assume that gates are applied sequentially.
% % (i.e., gate depth is equal to gate count)
%We refer to Sections~\ref{subsec:stateswghz} and~\ref{subsecstatesqft} for details about the gates in the W, GHZ and QFT circuits.
% \floris{Make sure these gates are explained or at least shown there.}
% \tim{gates are now in figure~\ref{fig:ghz-w-qft}}
 
\medskip
\noindent
$\rhd$ \emph{W state.} An $n$-qubit $W_n$ state preparation circuit has one $X$ gate, and $n-1$ controlled $G(p)$ gates and $n-1$ CNOT gates. As such, the overall number of gates is $\mathcal O(n)$.
% As the number of qubits $n$ increases, the computation cost of gates grows linearly, resulting in a complexity of \(\mathcal{O}\left(n\right)\).
\begin{itemize}[left=0pt]
\item For NumPy tensors, the $X$ gate is a single-qubit gate represented as a $2\times2$ matrix. 
Both the controlled $G(p)$ and the CNOT are two-qubit gates are represented as order-4 tensors of size $2\times2\times2\times2$. Applying each single- or two-qubit gate requires $\mathcal O(2^n)$ steps.
% because the gates are of constant size.
Given that there
are $\mathcal O(n)$ gates, the time complexity is $\mathcal{O}(n \cdot 2^n)$.
% \shih{It is correct that applying a quantum gate requires \(O(2^n)\) steps when the gates are of more than constant size because that's the complexity of matrix multiplication. }  

\item For the relational representation, we can see that the application of the $X$ gate keeps the number of non-zero elements invariant. In contrast, for the $W$-state preparation circuit, one can derive that that after the $k$-th controlled-NOT gate, the state is
\begin{equation}
\sqrt{\frac{1}{n}} \sum_{x\in F_k} \ket{x00\dots0} 
+ \sqrt{\frac{n-k}{n}}
\ket{
\underbrace{00\dots0}_{k\text{ zeroes}}100\dots0
}
    \label{eq:w-circuit-intermediate}
\end{equation}
where $F_k$ is the set of the first $k$ bitstrings $10\cdots0, 010\cdots00, \dots$ of length $k+1$ with hamming weight equal to 1.
This state has $k+1$ nonzero entries.
The controlled-$G(p)$ gate before each CNOT gate does not change that number.
Since applying a single- or two-qubit gate on state $\ket{\phi}$ takes at most $\mathcal O(\text{nnz}(\ket{\phi}))$ time, and since the $W$-state circuit has $n$ single- or two-qubit gates, we obtain a time complexity of $\mathcal{O}(n^2)$.
 % \emph{2)Relational representation.} Since the time complexity of the relational approach depends on the number of \texttt{JOIN}s executed. Due to the implementation the number of \texttt{JOIN}s is exactly the number of gates. Each \texttt{JOIN} itself will depend on the number of rows in each of the operands, since one operand will always be a gate tensor, we can consider its size constant as it's in the worst case for two-qubit gates $2^4=16$. The other operand is the stored state with its number of rows equal to the intermediate non-zero probability states. For the n-qubit W this leads to a time complexity of $\mathcal{O}(n^3)$.
\item
%\floris{This needs to be checked, not sure I understand it precise enough. In particular, if one applies a gate on an MPS, how does this affect bond dimension? And if one needs to do SVD to cut down increased bond, why is this then $\mathcal O(\chi^3)$ in the original bond dimension $\chi$ and not the new one?}
% \tim{Both Floris' comments have been addressed. For completness: bond dimension can square after two-qubit gate.}
For MPS, we recall from Sec.~\ref{subsec:MPS} that the time complexity for a single two-qubit gate is most $O(\chi^3)$.
As we fix $\chi$, and there are $n$ gates, an upper bound to the time complexity is $O(n \cdot \chi^3)$.
However, the intermediate states in the circuit as given in eq.~\eqref{eq:w-circuit-intermediate} have \emph{exact} bond dimension $2$, independently of $n$ or $k$. We thus conclude that the MPS time complexity for simulating the $W$ state circuit is $O(1)$ per gate, and as there are $O(n)$ gates in the circuit, the total time complexity is $O(n)$.

%, so that that total time complexity is constant 
%A careful analysis shows that the intermediate states in the W circuit
%Ideally, we would like to upper bound the time complexity by upper bounding $\chi_{\max}$. However, an upper bound for $\chi_{\max}$ for our case is 
% \floris{something is missing here?}
% can only bound the time complexity as it depends on the \tim{TODO rewrite}
%  that the $O(\chi)$ and $O(\chi^2)$ time complexities for a single- and two-qubit gate on adjacent qubits, respectively. %The bond dimension can square after two-qubit gates.
% If the gate spans $k$ qubits, the bond dimension can never grow more than $\chi \mapsto \chi^k$~\cite{perez2007matrix,vinkhuijzen2024knowledge}, although in practice the growth might be a lot smaller.
% Since 

% Since the QFT circuit contains two-qubit controlled gates between two qubits which have $n-2$ qubits in between, 
% Combining this with the gate count, the overall complexity is $\mathcal{O}(n\chi^3)$, where the $\chi$ in $\mathcal{O}(n \chi^3)$ is the largest bond dimension encountered during simulation. For the W state circuit, the bond dimension $\chi$ can be set to a constant value $\chi =2$, reducing the final complexity to $\mathcal{O}(n)$. 
\end{itemize}
% \begin{figure}[h]
% \centering
%  \includegraphics[width=0.8\linewidth]{figures/MPS_gates.jpg}  
 
% \caption{Applying gates to qubits in MPS}
% \label{fig:mps_gates}
%  % \vspace{-0.6cm}
% \end{figure}

\medskip
\noindent
$\rhd$\emph{GHZ state.} For the $n$-qubit GHZ$_n$ state circuit, we note that after the $k$-th CNOT, the state is 
\[
\frac{1}{\sqrt{2}}\ket{00\dots0} + 
\frac{1}{\sqrt{2}}\ket{\underbrace{11\dots1}_{k+1 \textnormal{ times}}00\dots0}
\]
which has exact bond dimension $2$, making its MPS time complexity similar to that of the $W$ state.
Since this intermediate state has only $2$ nonzero entries, independently of $k$ or $n$, updating the relational encoding upon applying a CNOT is done in constant time, making the time complexity $O(n)$ as there are $O(n)$ gates.
The time complexity analysis for the NumPy tensor format is similar to the $W_n$ state. %Instead of two controlled gates in the GHZ state preparation circuit there are exactly $n$ gates, starting with one Hadamard gate on the first qubit followed by $n-1$ CNOT gates.
%Therefore the time complexity only differs in a constant factor from the $W$ state preparation circuit.
%For relational representation, the reasoning is similar to the $W_n$ state preparation circuit, which results in a linear time complexity, except that the number of non-zero states remains constant throughout the computation (two). This results in an overall time complexity of $\mathcal O(n^2)$.

\medskip
\noindent
$\rhd$\emph{QFT.} The QFT circuit consists of $\mathcal{O}(n^2)$ single- and two-qubit gates, so the time complexity for NumPy tensors becomes $\mathcal{O}(n^2 \cdot 2^n)$.
% \shih{there are implementations using $n(n+1)/2$ gates. I think it would be safe to say the circuit consists of $\mathcal{O}(n^2)$ gates.} 
%The analysis for NumPy  tensors is analoguous to $W_n$ state, i.e., 
For the relational representation, the time complexity scales similar to the NumPy tensor approach except for an additional factor $n$ for the indices. That is, its time complexity is $\mathcal{O}(n^3 \cdot 2^n)$.
For MPS, a single-qubit gate update runs in time $O(\chi^3)$, and so does the two-qubit gate as long as the qubits are adjacent.
If this is not the case, as in the QFT circuit, additional SWAP gates are needed to bring the two qubits next to each other.
For the QFT circuit, at most $O(n)$ SWAP gates are needed to make each gate act on adjacent qubits, increasing the gate count to $O(n^3)$, and hence the time complexity to $O(n^3 \chi^3)$.
We expect that $\chi_{\text{exact}}$ is constant for the $\ket{QFT_n}$ state, which might improve the bound to $O(n^3)$ when the bond dimension $\chi$ used in simulation is chosen $\chi \leq \chi_{\text{exact}}$, but we consider computing it exactly out of scope for this  analysis.

%depends on the initial state as this would influence the intermediate states. In our experiments, we start with the initial state of all qubits in state $\ket{0}$, hence there is no entanglement in the system. 
%\floris{Again someone with more expertise in MPS check the following.}
%Therefore all intermediate states can be represented with maximum accuracy using a maximum bond dimension of $2$. In general the bond dimension will influence the size of the matrices in the MPS and hence the time complexity. With the number of gates being $n^2$ this leads to a time complexity of $\mathcal{O}(n^2 \cdot \chi^3)$.

% \shih{took a pass on B2; will double check MPS for GHZ and QFT tomorrow (Dec. 28) by doing calculations}

% {\shih{In Fig. 2, I personally prefer $\ket{+}^{\otimes n}$ to $\ket{\mathrm{QFT}_n}$ for a uniform superposition but it's just notation}}

\subsection{Experiment setting}
\label{ssec:setting}
In this section, we explain our experimental setting. We report our preliminary results in Sec.~\ref{ssec:results}.
% \begin{enumerate}
%     \item What advantages of relational databases could impact quantum circuit simulations
%     \item What type of circuits could benefit most from using relational databases
%     \item What optimizations could be done on the data management / database systems
% \end{enumerate}

% \para{Hardware} 
We conducted our experiments on  on a machine equipped with an Intel Core i9-12900K CPU (16 cores, up to 5.2 GHz), 64 GB RAM, and running Ubuntu 22.04 LTS. Each experiment was executed on a single CPU core, and we report the average execution time across 100 runs.
% \para{Software} 

\para{Baselines and RDBMS implementation} We have implemented two NumPy-based baselines with different data representations: 
\begin{itemize}[left=0pt]
\item \emph{Baseline I}: tensor network representation and tensor contractions are performed over order-$n$ tensor, where $n$ is the number of qubits. This Baseline used the NumPy tensor format described in Section~\ref{sec:ana}. We have chosen this baseline as it is a general approach.  
% employs the same data representation (order-n tensor)
\item \emph{Baseline II}: Matrix Product State (MPS) representation, where again each component is stored in NumPy format (see again Section~\ref{sec:ana}). We have chosen this baseline as MPS is widely used and often considered as one of the state-of-the-art data representations for classical simulation. 
\end{itemize}
Both baselines were developed in Python 3.13.0 and NumPy 2.1.3, widely used linear algebra libraries for classical simulation.
\begin{itemize}[left=0pt]
\item \emph{RDBMS-based solution}: we have implemented and evaluated the relational representation in Sec.~\ref{sec:rel} with three database systems: PostgreSQL 17.2, SQLite 2.6.0, DuckDB 1.1.3. 
The tested relational databases include a PostgreSQL database running in a docker container, DuckDB used in-memory via the Python API and an in-memory instance of a SQLite database. 
\end{itemize}
% In all cases we measure only the execution time of the tensor contractions and in case of the database computations the conversion of Einstein summation notation to SQL query. Excluded from the runtime are converting the circuit into Einstein Summation, preparing the ideal contraction path and connecting to the database.\\ 

\noindent\textbf{ Why these baselines?} The state-of-the-art approach in \cite{einsteinSQL2023} treats quantum states as order-n tensors and transforms the simulation circuits into a series of SQL queries executed in relational databases. In Sec.~\ref{sec:rel}, we further refine this RDBMS-based solution by explicitly defining its relational representation. We select Baseline I because it also adopts an order-n tensor model, enabling a direct comparison of whether the sparse relational encoding in an RDBMS delivers any tangible benefits. We choose Baseline II (MPS) because Matrix Product States are widely acknowledged in the quantum computing community as a more sophisticated and frequently used representation for qubit states. Including MPS allows us to evaluate how well the relational representation competes with this state-of-the-art approach.

As an initial effort, our preliminary findings emphasize representation rather than optimization. As discussed in Sec.~\ref{sssec:opt}, existing research prototypes and commercial simulators often employ parallelization, SIMD processing, and matrix decomposition, which substantially influence efficiency. Addressing Q1, i.e., whether it is advantageous to push simulations into databases or whether a database can exploit memory more effectively than established simulators, remains an open avenue for future research. In particular, leveraging native database features such as indexing, caching, and exploring the emerging tensor-based database technologies mentioned in Q2 will require more system-building efforts to determine how database technologies might be enhanced for large-scale quantum circuit simulations.

\para{Input circuit generation} The inputs of a classical simulator are quantum circuits. In preliminary experiments, we evaluated the baselines and RDBMS solutions over three widely used circuits: W state preparation circuit, GHZ state preparation circuit and QFT circuit. 
We have developed a circuit generator which is made publicly available online.\footnote{\url{https://github.com/infiniData-Lab/Quantum}}
The input circuits are generated in JSON-format with a given number of qubits and a list of gates which are JSON-objects themselves containing an identifier like "H" for the Hadamard gate and a list of the qubits it acts on. The ideal contraction path is part of the input and generated by the Python library opt\_einsum 3.3.0\footnote{\url{https://pypi.org/project/opt-einsum/}} providing the generated Einstein summation notation in the format "ij,jk->ik". This notation corresponds to a tensor contraction described by the Einstein summation notation $C_{ik} = \sum_j A_{ij}B_{jk}$, where $i,j,k$ are the indices of the corresponding tensors $A$, $B$ and $C$. 
% See also Section~\ref{subsec:tensors}.

\begin{figure*}[t]
    \centering
    \begin{subfigure}[b]{0.8\linewidth} % Adjusted width to fit three subfigures in one row
        \centering
        \includegraphics[width=0.8\linewidth]{figures/experiments/2212/std_legend.png}
        \phantomsubcaption
        \label{fig:mem_legend_app}
     \end{subfigure}
     \setcounter{subfigure}{0}
    \begin{subfigure}[b]{0.32\linewidth} % Adjusted width to fit three subfigures in one row
        \centering
        \includegraphics[width=\linewidth]{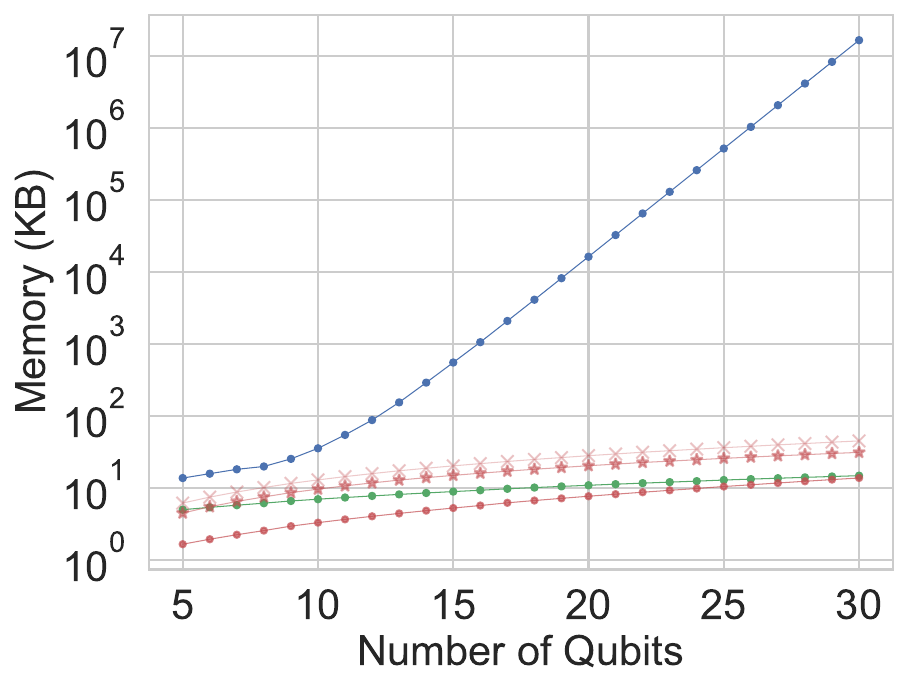}
        \caption{W state (\textbf{sparse} circuit)}
        \label{fig:wstate_mem}
     \end{subfigure}
     \hfill
     \begin{subfigure}[b]{0.32\linewidth} % Adjusted width to fit three subfigures in one row
        \centering
        \includegraphics[width=\linewidth]{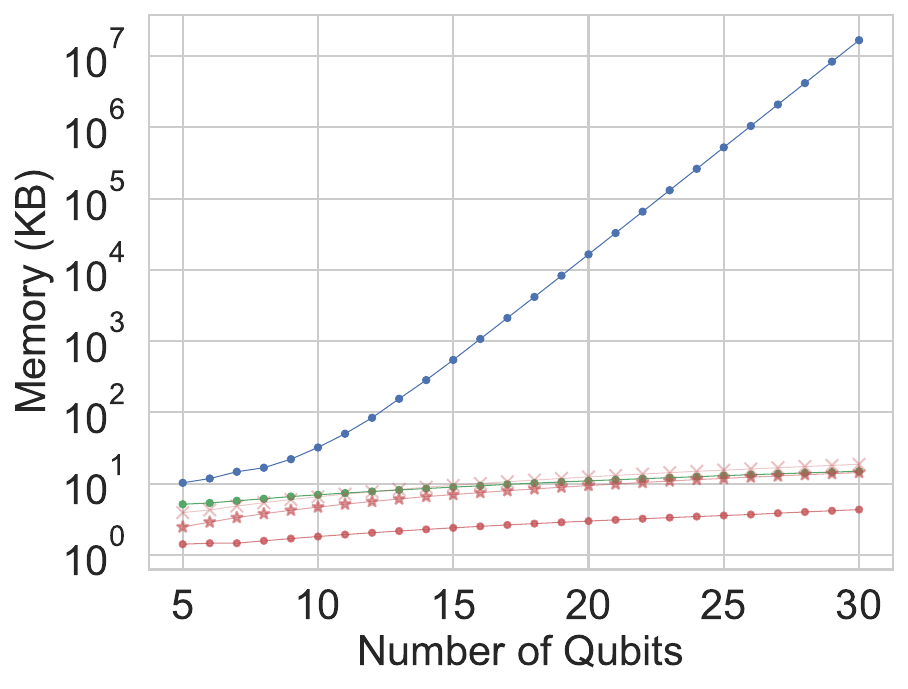}
        \caption{GHZ state (\textbf{sparse} circuit)}
        \label{fig:ghzstate_mem}
     \end{subfigure}
    \hfill
      \begin{subfigure}[b]{0.32\linewidth} % Adjusted width to fit three subfigures in one row
        \centering
        \includegraphics[width=\linewidth]{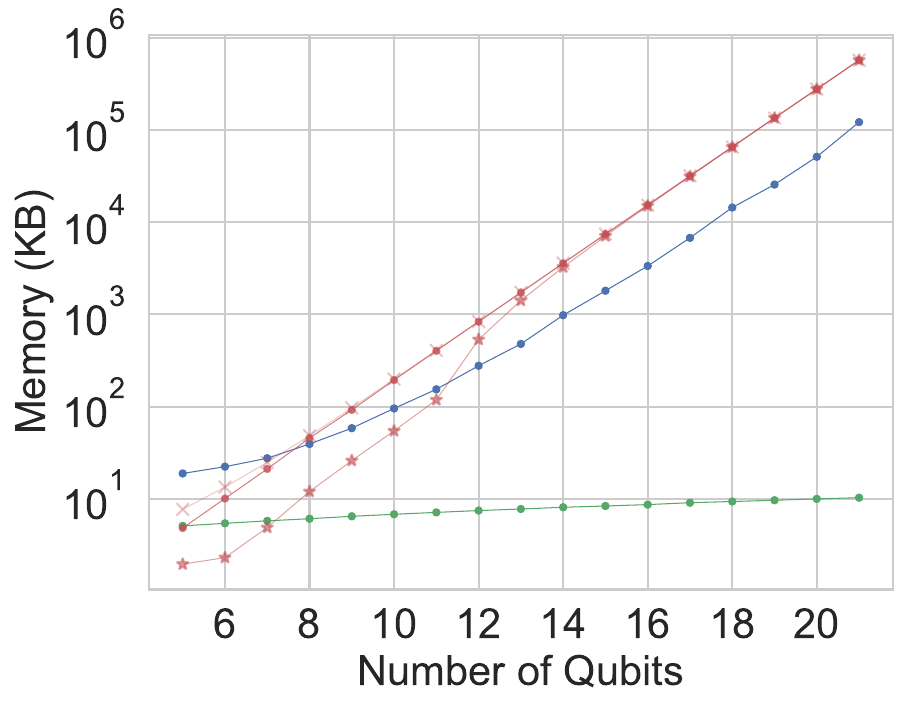}
        \caption{QFT (\textbf{dense} circuit)}
        \label{fig:mem_qft}
     \end{subfigure}
         \caption{Memory usage comparison of three RDBMS-based solutions (red) against general order-n tensor baseline (blue) and SotA MPS baseline (green). The x-axis shows increasing numbers of qubits, and the y-axis shows memory consumption (KB) in $\log_{10}$-scale. For sparse circuits (W and GHZ), RDBMS solutions demonstrate significantly lower memory usage than the order-n tensor baseline, with SQLite showing a slight advantage over the MPS baseline. For the dense circuit QFT, the memory usage of RDBMS solutions grows comparably to or exceeds the order-n tensor baseline, indicating the need for improvements in handling dense tensor computations.}
    \label{fig:mem}\Description[]{}
\end{figure*}

\begin{figure}[t]
    \centering
    \includegraphics[width=0.7\linewidth]{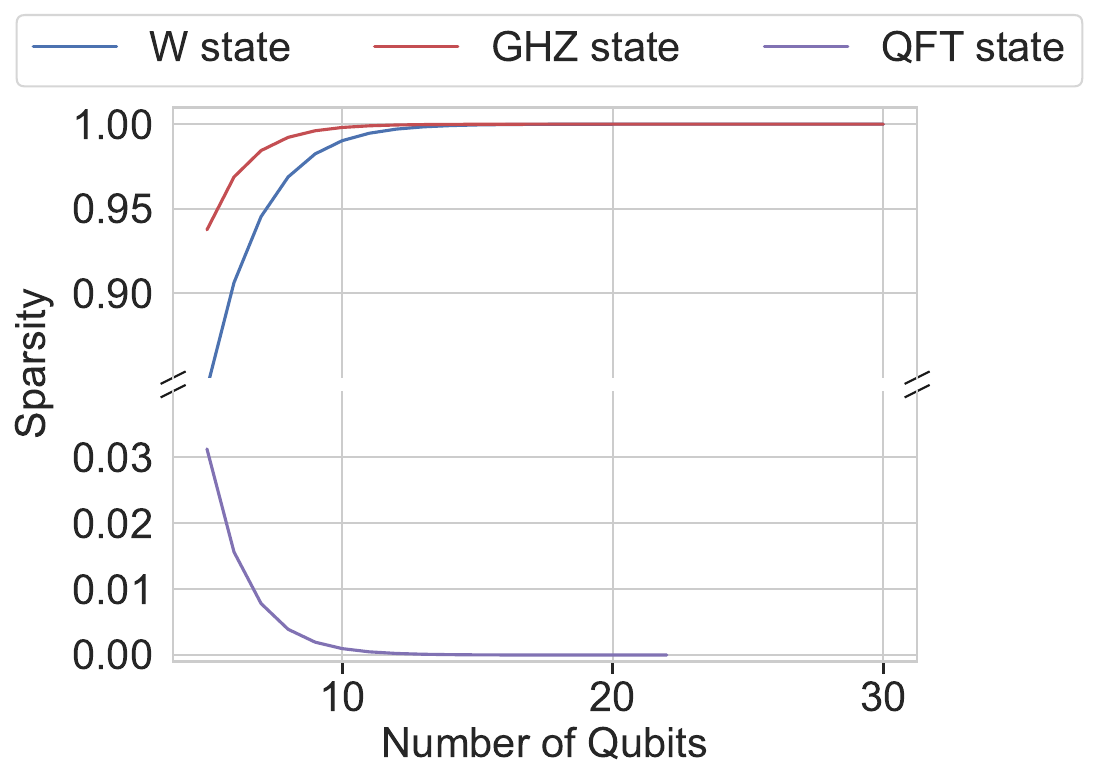}
    \caption{Sparsity ($s=1-\frac{\text{nnz}(\ket{\Psi})}{2^n}$) comparison for GHZ and W state preparation circuits and QFT circuit, where $\text{nnz}(\ket{\Psi})$ denotes the number of  non-zero probability amplitudes in the state $\ket{\Psi}$, $n$ is the number of qubits and $2^n$ is total number of possible states.}
       \label{fig:sparsity_GW}\Description[]{}
\end{figure}

For RDBMS solutions, our circuit generator produces SQL queries representing the circuits,  following the methodology in state-of-the-art for transforming quantum circuits into SQL queries \cite{einsteinSQL2023}. 
%
% These circuits are the circuit which generates the n-qubit Greenberger–Horne–Zeilinger state i.e. $\frac{1}{\sqrt{2}}(\ket{00..0} + \ket{11..1})$, the circuit which generates the n-qubit W state i.e. $\frac{1}{\sqrt{n}}(\ket{00..01} + \ket{00..10} + .. + \ket{10..00})$, n-qubit Quantum Fourier Transform circuit and n-qubit Quantum Phase Estimation circuit for a simple unitary matrix that is equivalent to the Pauli-Z-Gate.\\
% \para{SQL query generation}  
The generated SQL queries include the representation of the initial quantum state — typically the zero state, represented as the vector $\ket{0}$ — and the quantum gates as tensors. The tensor contraction paths of input circuit are specified by the Einstein summation notation, which are translated into SQL operations. 
Specifically, tensor contractions are expressed as \texttt{SUM} operations over the shared index in SQL, with tensors as relational tables. The resulting tensor is grouped by the output indices.
The contraction order is optimized using Common Table Expressions (CTEs), enabling efficient computation and reuse of intermediate results. An example of applying a Hadamard gate followed by a $CNOT$-gate to a two-qubit system would look as follows with \texttt{T0}, \texttt{T1} and \texttt{T2} as the intermediate states and \texttt{CX} and \texttt{H} as the corresponding gates:
\begin{verbatim}
WITH T0(s0, r) AS (
      VALUES 
      (0, 1.0)),
     H(i, j, r) AS (
      VALUES 
      (0, 0, 0.7071067811865475), 
      (0, 1, 0.7071067811865475), 
      (1, 0, 0.7071067811865475), 
      (1, 1, -0.7071067811865475)),
     CX(i, j, k, l, r) AS (
      VALUES 
      (0, 0, 0, 0, 1.0), 
      (0, 1, 0, 1, 1.0), 
      (1, 0, 1, 1, 1.0), 
      (1, 1, 1, 0, 1.0)),
  T1 AS (
  SELECT CX.i AS i, CX.j AS s1, CX.k AS s0, 
  SUM(CX.r * T0.r) AS r FROM CX, T0 WHERE CX.l=T0.s0 
  GROUP BY CX.i, CX.j, CX.k),
  T2 AS (
  SELECT H.i AS s1, 
  SUM(H.r * T0.r) AS r FROM H, T0 WHERE H.j=T0.s0 
  GROUP BY H.i) SELECT T1.s1 AS s1, T1.s0 AS s0, 
  SUM(T2.r * T1.r) AS r FROM T2, T1 WHERE T2.s1=T1.s1 
  GROUP BY T1.s1, T1.s0 ORDER BY s1, s0
\end{verbatim}

\subsection{Preliminary results}
\label{ssec:results}

In Sec.~\ref{ssec:dbq2}, we have posed the first research question (Q1): Should simulation workloads be pushed to existing DBMSs? 
Based on the complexity analysis in Tables 1 and 2, RDBMS solutions appear to offer advantages over order-n tensor representation, when input circuits involve sparse tensor computation, such as W state and GHZ state preparation circuits. To investigate this, we compare RDBMS solutions with baselines in terms of memory usage (Sec.~\ref{sssec:memory}) and runtime (Sec.~\ref{sssec:time}).

Furthermore, in Sec.~\ref{ssec:q_gen}, through preliminary experiments, we have discovered a research gap in the state-of-the-art \cite{einsteinSQL2023}: the inefficiency of SQL query generation for simulation workloads. Understanding this gap is crucial for future work toward developing an end-to-end simulation pipeline based on database systems.

% For each implemented circuit we evaluate the memory usage and execution time. Circuits that generate dense states are represented in logarithmic scale, as memory usage and runtime often scale exponentially with the number of qubits. While the number of qubits serves as the primary parameter, we also assess the impact of limited memory availability on the simulations.

 %=======================================================================%
\begin{figure*}[t]
    \centering
    \begin{subfigure}[b]{0.32\linewidth} % Adjusted width to fit three subfigures in one row
        \centering
        \includegraphics[width=\linewidth]{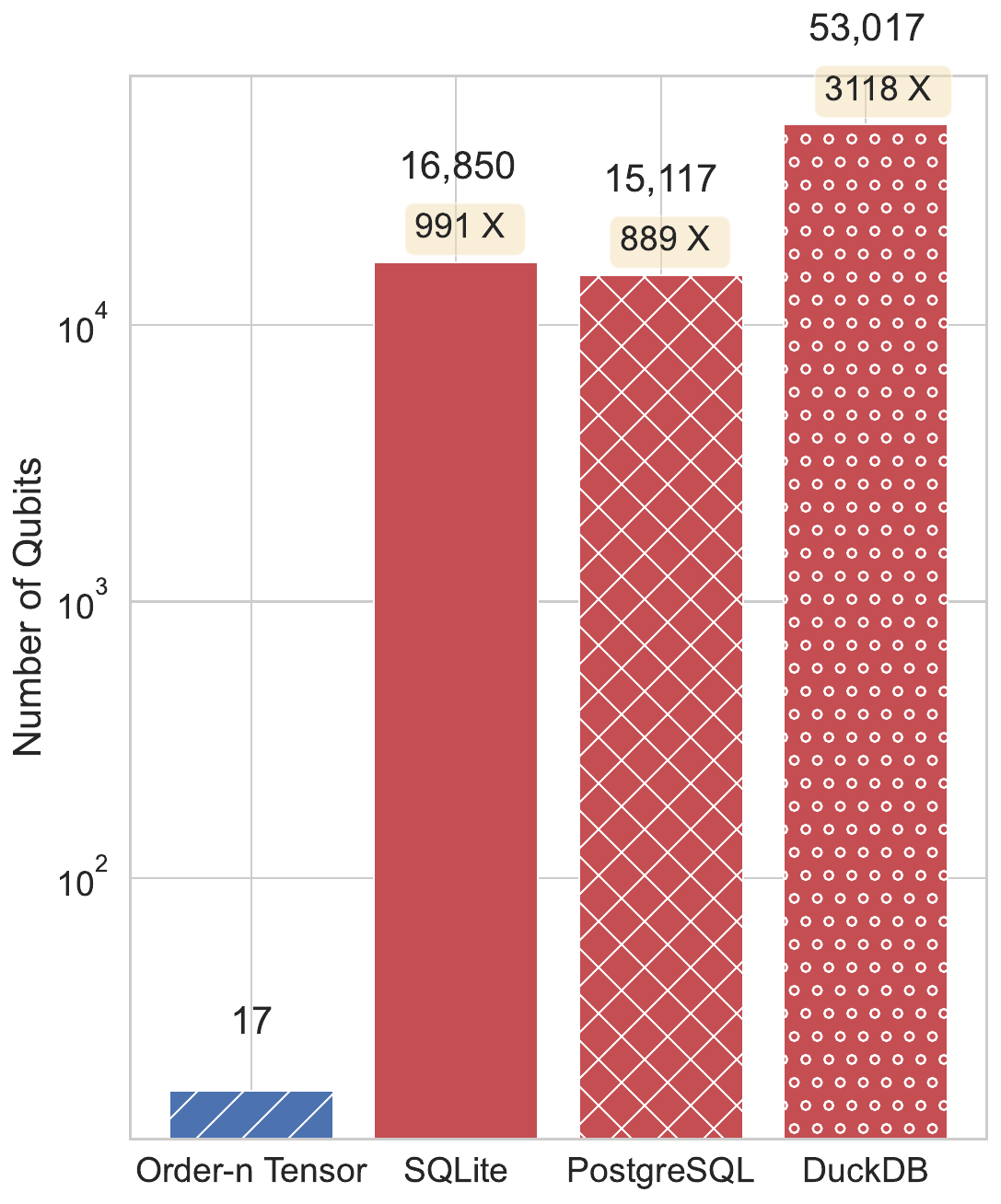}
        \caption{W state (\textbf{sparse} circuit)}
        \label{fig:memlimit_wstate}
     \end{subfigure}
     \hfill
     \begin{subfigure}[b]{0.32\linewidth} % Adjusted width to fit three subfigures in one row
        \centering
        \includegraphics[width=\linewidth]{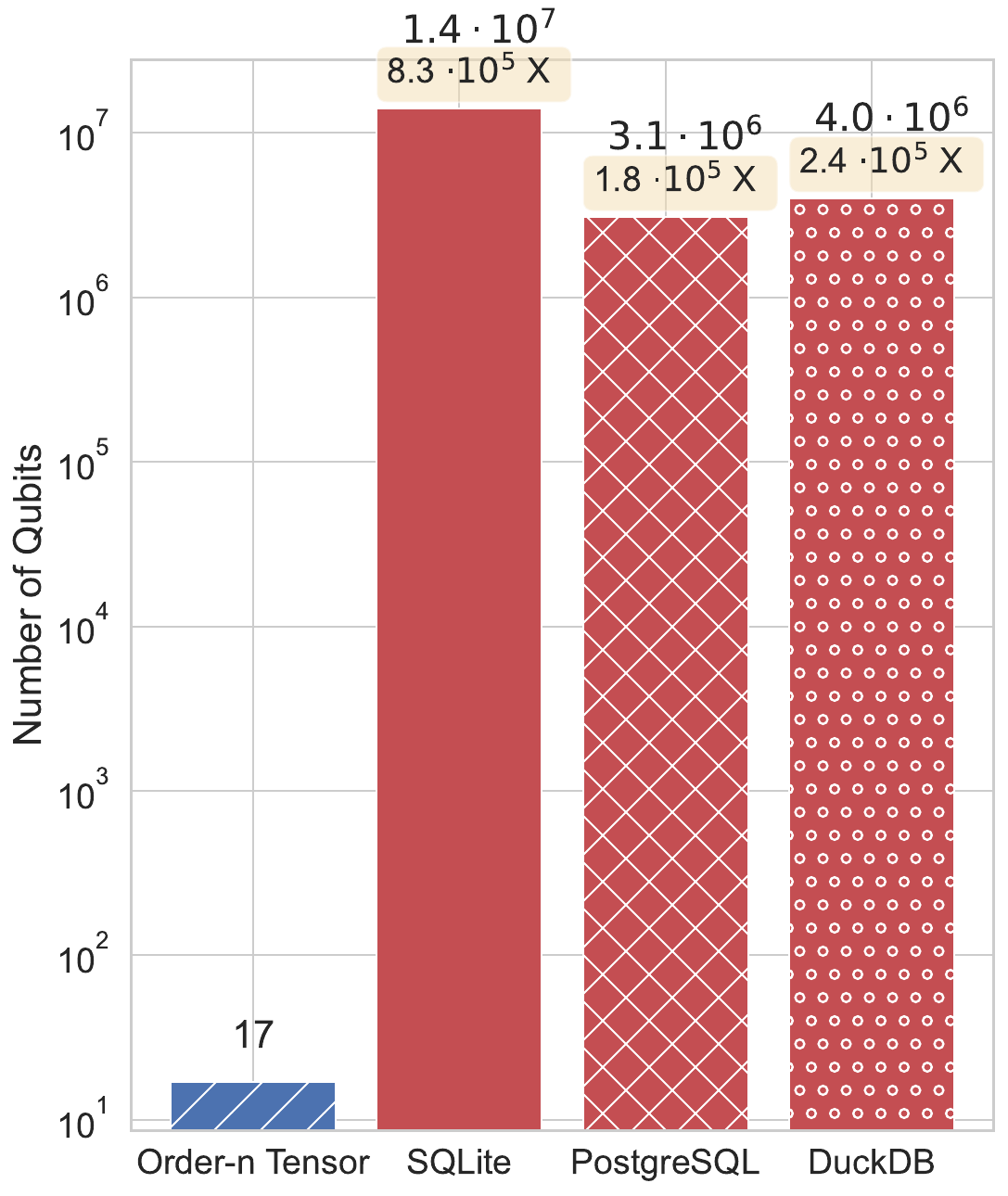}
        \caption{GHZ state (\textbf{sparse} circuit)}
        \label{fig:memlimit_ghz}
     \end{subfigure}
    \hfill
      \begin{subfigure}[b]{0.32\linewidth} % Adjusted width to fit three subfigures in one row
        \centering
        \includegraphics[width=\linewidth]{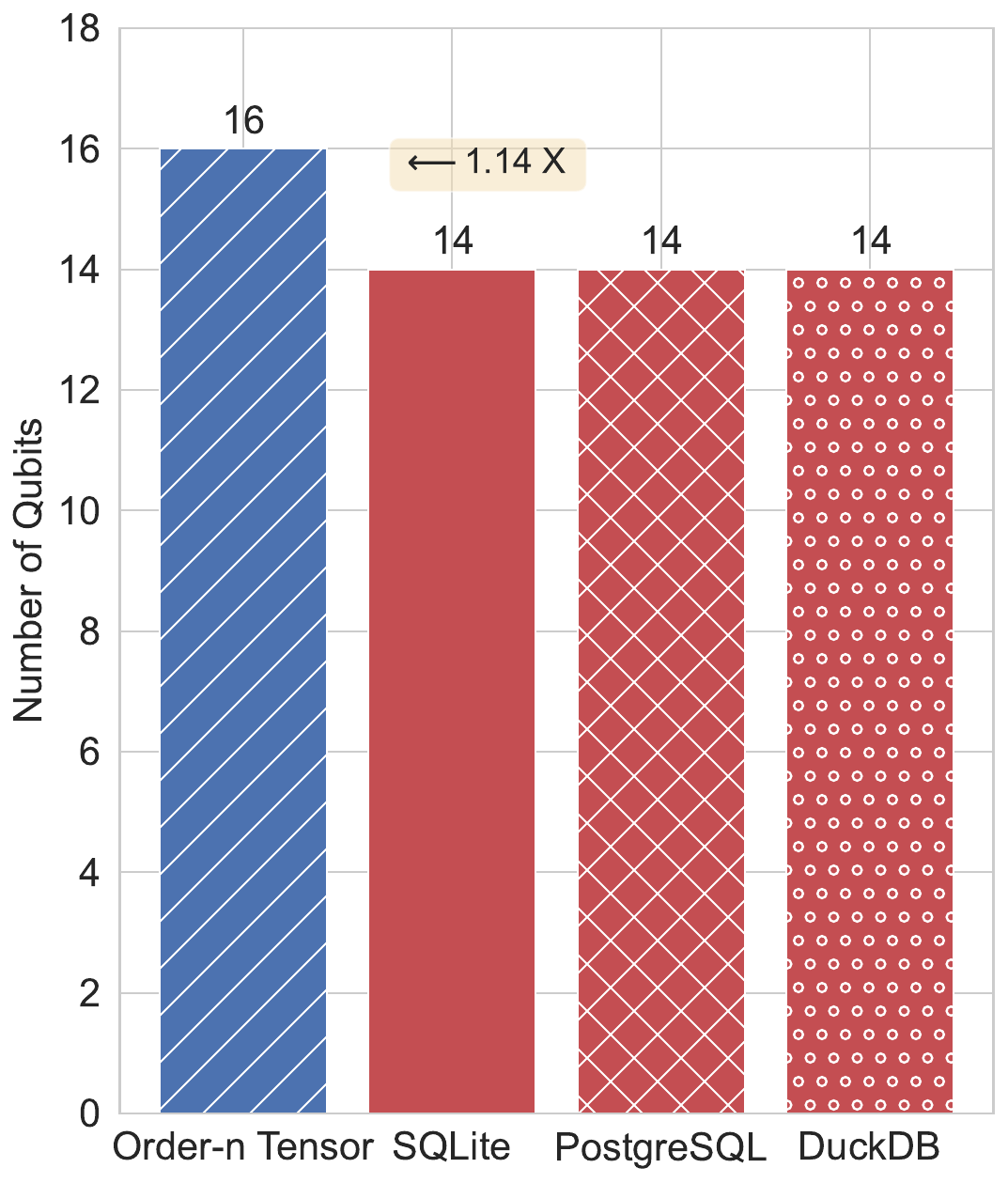}
        \caption{QFT (\textbf{dense} circuit)}
        \label{fig:memlimit_qft}
     \end{subfigure}
         \caption{Comparison of the simulated numbers of qubits for RDBMS solutions and the order-n baseline under a 2.0 GB memory limit. The y-axis is in log\(_{10}\)-scale for W and GHZ state preparation circuits. RDBMS-based approaches exhibit significant scalability in sparse circuits, simulating up to 16,850, 15,117, and 53,017 qubits on average for W states---yielding improvements of $991\times$, $889\times$, and $3,118\times$, respectively. For GHZ states, RDBMS solutions simulate $1.4 \cdot 10^7$, $3.1 \cdot 10^6$, and $4.0 \cdot 10^6$ qubits, achieving enhancements of $8.3 \cdot 10^5\times$, $1.8 \cdot 10^5\times$, and $2.4 \cdot 10^5\times$, respectively. For the dense circuit QFT, RDBMS solutions simulate up to 14 qubits, slightly below the 16 qubits supported by the order-\(n\) tensor baseline, indicating the need for further improvements. 
  }
    \label{fig:memlimit}\Description[]{}
\end{figure*}
\begin{figure*}[t]
    \centering
    \begin{subfigure}[b]{0.8\linewidth} % Adjusted width to fit three subfigures in one row
        \centering
        \includegraphics[width=0.8\linewidth]{figures/experiments/2212/std_legend.png}
        \phantomsubcaption
        \label{fig:mem_legend_two}
     \end{subfigure}
    \setcounter{subfigure}{0}
    \begin{subfigure}[b]{0.32\linewidth} % Adjusted width to fit three subfigures
        \centering
        \includegraphics[width=\linewidth]{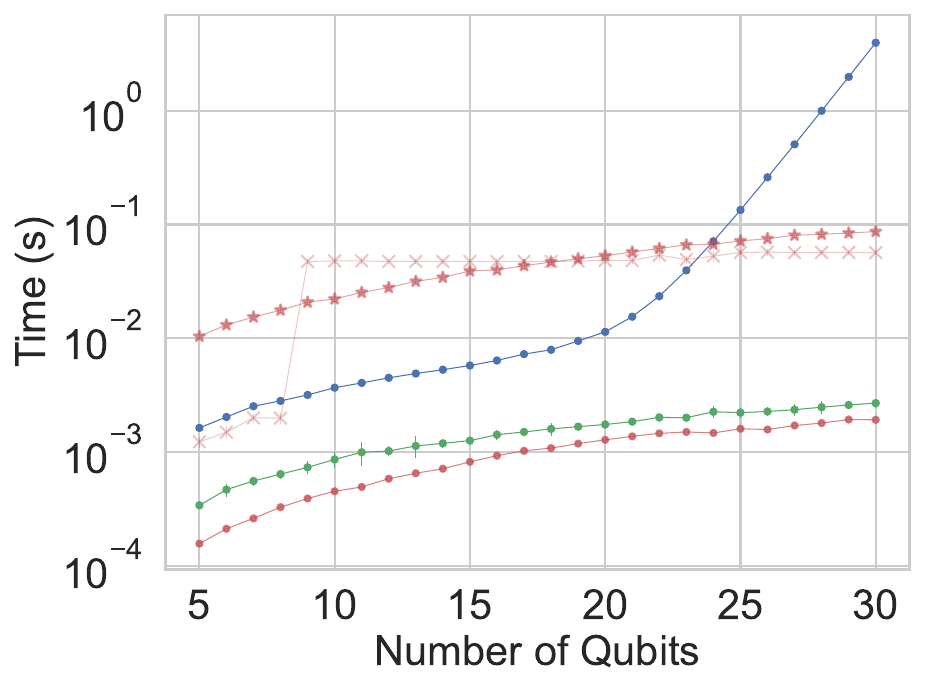}
        \caption{W state (\textbf{sparse} circuit)}
        \label{fig:wstate_time}
     \end{subfigure}
     \hfill
     \begin{subfigure}[b]{0.32\linewidth} % Adjusted width to fit three subfigures
        \centering
        \includegraphics[width=\linewidth]{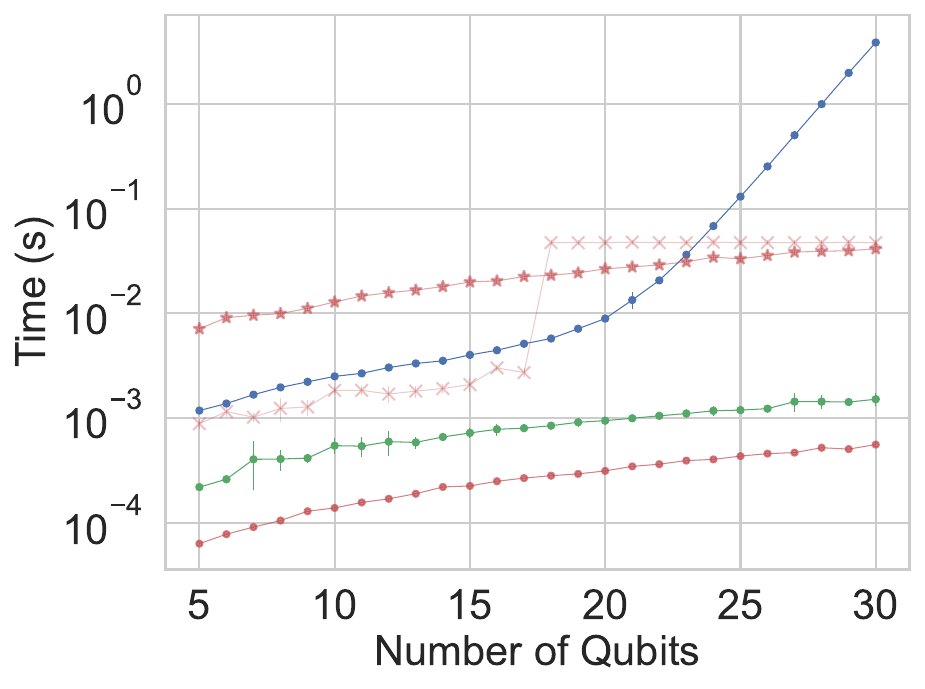}
        \caption{GHZ (\textbf{sparse} circuit)}
        \label{fig:ghz_time}
     \end{subfigure}
      \hfill
      \begin{subfigure}[b]{0.32\linewidth} % Adjusted width to fit three subfigures
        \centering
        \includegraphics[width=\linewidth]{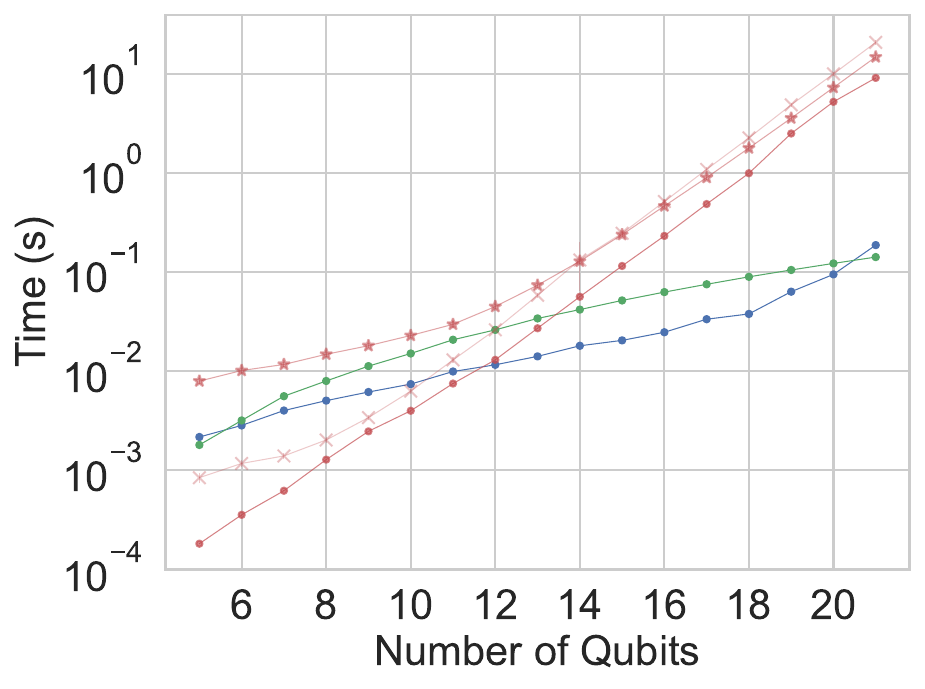}
        \caption{QFT (\textbf{dense} circuit)}
        \label{fig:qft_time}
     \end{subfigure}
       \caption{Runtime comparison of three RDBMS-based solutions (red) against general order-n tensor baseline (blue) and SotA MPS baseline (green). The y-axis is in log\(_{10}\) scale. The SQLite solution outperforms both baselines for sparse circuits. For the dense circuit QFT, NumPy-based baselines handle dense tensor computations more efficiently, indicating the need for further optimizations in RDBMS solutions.
  }
    \label{fig:exp}\Description[]{}
\end{figure*}
% \begin{figure}[t]
%     \centering
%     \begin{subfigure}[b]{0.46\linewidth}
%         \centering
%         \includegraphics[width=\linewidth]{figures/experiments/0512/W_circuit_time_log_w_MPS.png}
%         \caption{W state (\textbf{sparse} circuit)}
%         \label{fig:wstate_time}
%      \end{subfigure}
%      \hfill
%       \begin{subfigure}[b]{0.46\linewidth}
%         \centering
%         \includegraphics[width=\linewidth]{figures/experiments/0512/QFT_circuit_memory_log_w_MPS.png}
%         \caption{QFT (\textbf{dense} circuit)}
%         \label{fig:qft_time}
%      \end{subfigure}
%     \caption{Execution time comparison of RDBMS-based vs. NumPy-based solutions}%: $a)$  circuit GHZ $b)$ Dense circuit QFT}
%     \label{fig:exp}
%     % \vspace{-6mm}
% \end{figure}
%=======================================================================%

\vspace{0.2cm}
\begin{myframe}[roundcorner=3pt,innerleftmargin = 5pt]
{\bf Q1.} \textit{Should we push the simulation workload to existing DBMSs?  
\begin{enumerate}[label=(\alph*)]
 \item When to use the relational representation?
\end{enumerate}
 }
\end{myframe}

% \vspace{0.2cm}
% \begin{myframe}[roundcorner=3pt,innerleftmargin = 5pt]
% {\bf 1} \textit{When to use relational representation?} 
% \end{myframe}

\subsubsection{Memory Usage} \label{sssec:memory}Fig.~\ref{fig:mem} compares the memory usage %(peak memory allocation) 
of RDBMS solutions and two baselines.\footnote{In all QFT-related experiments in this section, we were limited to 21 qubits due to the substantial time required to complete 100 runs, and the memory limit of our hardware.}
% The memory usage of each simulation was analyzed by tracking the peak memory allocation during tensor contractions. 
The memory usage of the order-n tensor baseline grows exponentially, consistent with its $\mathcal{O}(2^n)$ memory complexity in Table~\ref{tab:representation}. This rapid growth becomes evident as the number of qubits increases. 
In contrast, the RDBMS solutions, i.e., SQLite, PostgreSQL, and DuckDB, demonstrate significantly lower memory consumption in sparse scenarios (W state and GHZ state preparation circuits).  
% RDBMS solutions have exhibited low memory usage, with small variations as the number of qubits increases. 
Memory usage for RDBMS solutions exhibits small variation with increasing qubits, aligning with the space complexity analysis in Table \ref{tab:representation}, which predicts a linear growth for GHZ states and a quadratic growth for W states. 
This memory efficiency arises from the RDBMS solutions' ability to compactly represent non-zero values, providing an effective form of sparse tensors in simulation.

%----------------------------------------------------------%
\begin{figure*}[t]
    \centering
    \begin{subfigure}[b]{0.8\linewidth} % Adjusted width to fit three subfigures in one row
        \centering
        \includegraphics[width=0.4\linewidth]{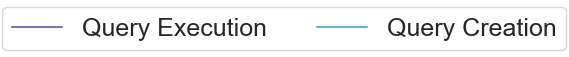}
        \phantomsubcaption
        \label{fig:query_mem_legend}
     \end{subfigure}
    \setcounter{subfigure}{0}
    \begin{subfigure}[b]{0.32\linewidth} % Adjusted width to fit three subfigures
        \centering
        \includegraphics[width=\linewidth]{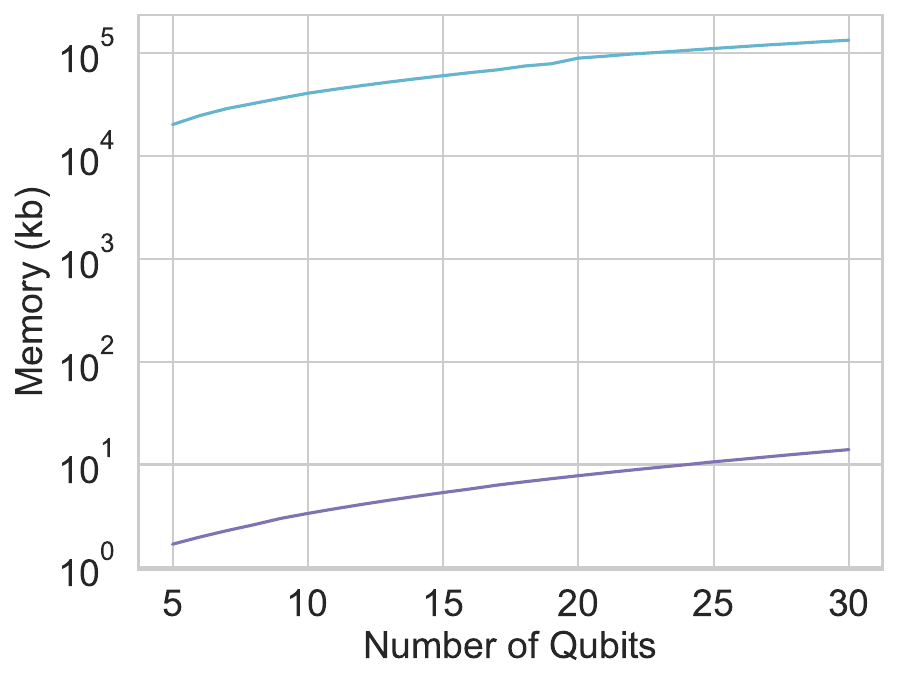}
        \caption{W state (\textbf{sparse} circuit)}
        \label{fig:wstate_memusg}
     \end{subfigure}
     \hfill
     \begin{subfigure}[b]{0.32\linewidth} % Adjusted width to fit three subfigures
        \centering
        \includegraphics[width=\linewidth]{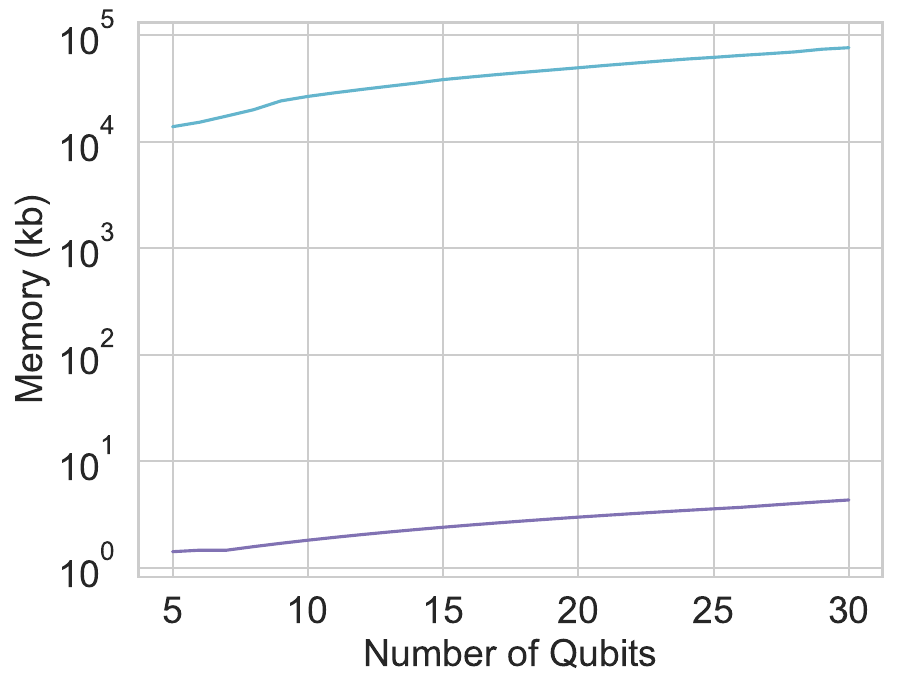}
        \caption{GHZ state (\textbf{sparse} circuit)}
        \label{fig:ghz_memusg}
     \end{subfigure}
     \hfill
      \begin{subfigure}[b]{0.32\linewidth} % Adjusted width to fit three subfigures
        \centering
        \includegraphics[width=\linewidth]{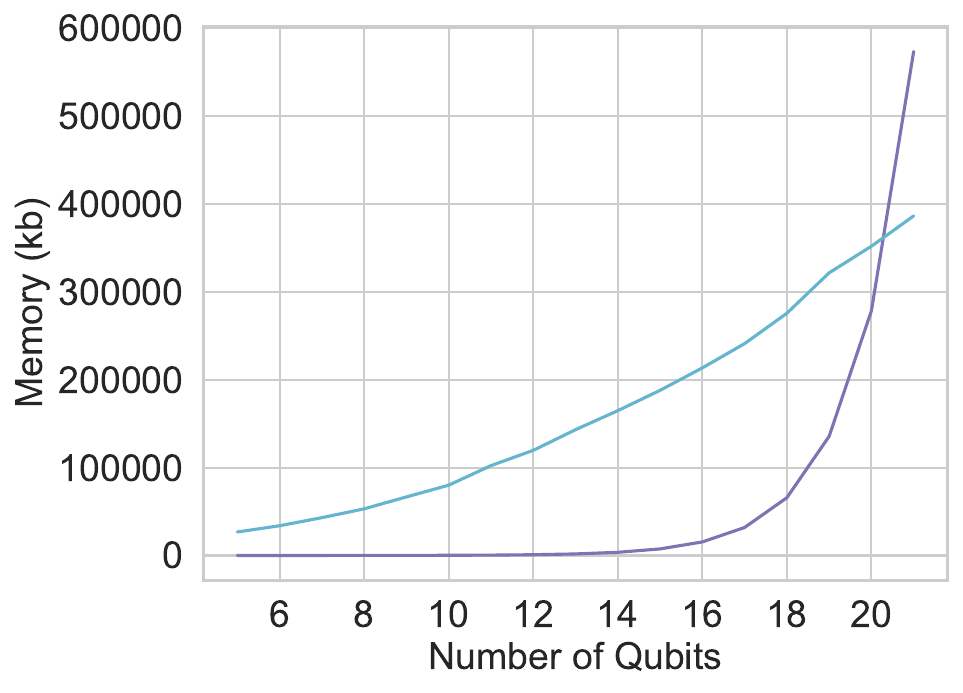}
        \caption{QFT (\textbf{dense} circuit)}
        \label{fig:qft_memusg}
     \end{subfigure}
    \caption{Memory usage comparison of SQL query generation \cite{einsteinSQL2023} vs. SQL query execution for SQLite. The y-axis in (a) and (b) uses $\log_{10}$-scale due to significant memory usage differences. Query generation incurs substantial memory overhead, in sharp contrast to query execution, especially for sparse circuits like W and GHZ states, where RDBMS solutions are more memory-efficient. This overhead results from tensor transformations and the translation of contraction paths into SQL join operations in \cite{einsteinSQL2023}.
  }
    \label{fig:mem_qcreate}\Description[]{}
\end{figure*}
\begin{figure*}[t]
    \centering
    \begin{subfigure}[b]{0.8\linewidth} % Adjusted width to fit three subfigures in one row
        \centering
        \includegraphics[width=0.4\linewidth]{figures/experiments/2212/query_legend.png}
        \phantomsubcaption
        \label{fig:query_mem_legend_gen}
     \end{subfigure}
    \setcounter{subfigure}{0}
    \begin{subfigure}[b]{0.32\linewidth} % Adjusted width to fit three subfigures
        \centering
        \includegraphics[width=\linewidth]{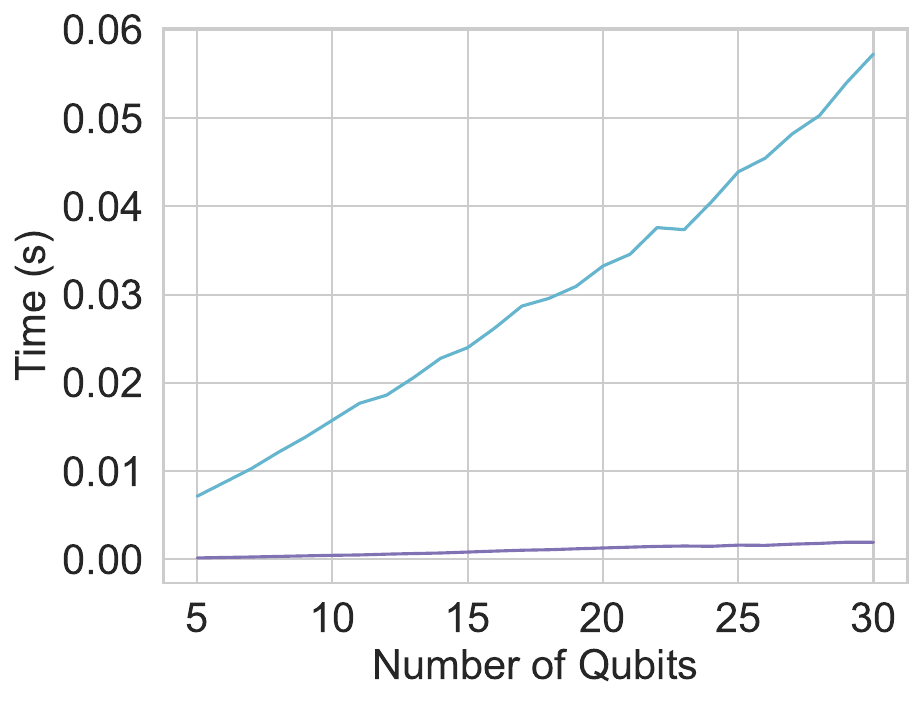}
        \caption{W state (\textbf{sparse} circuit)}
        \label{fig:wstate_extime}
     \end{subfigure}
     \hfill
     \begin{subfigure}[b]{0.32\linewidth} % Adjusted width to fit three subfigures
        \centering
        \includegraphics[width=\linewidth]{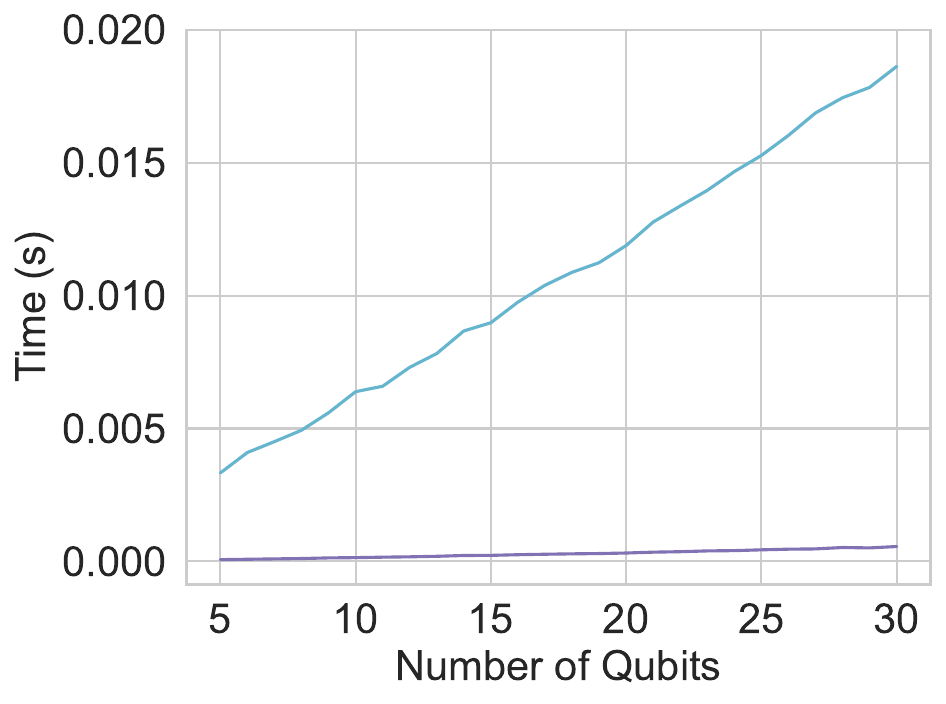}
        \caption{GHZ state (\textbf{sparse} circuit)}
        \label{fig:ghz_extime}
     \end{subfigure}
     \hfill
      \begin{subfigure}[b]{0.32\linewidth} % Adjusted width to fit three subfigures
        \centering
        \includegraphics[width=\linewidth]{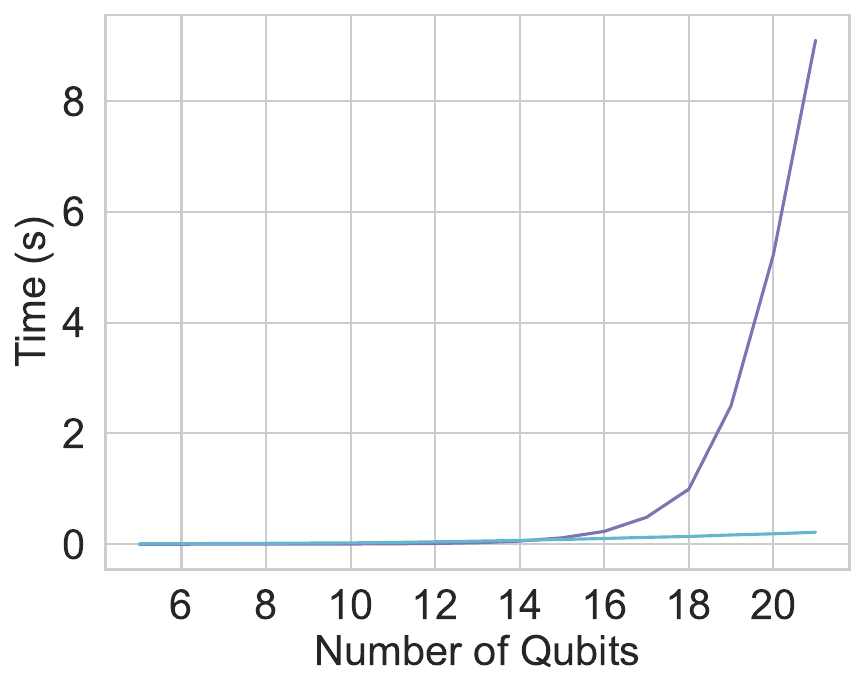}
        \caption{QFT (\textbf{dense} circuit)}
        \label{fig:qft_extime}
     \end{subfigure}
    \caption{Comparison of SQL query generation time \cite{einsteinSQL2023} vs. SQL query execution time for SQLite.  Query generation requires more time than execution for sparse circuits like W and GHZ states, where RDBMS solutions are more efficient. Query generation incurs higher time costs due to the string construction proportional to the number of gates, along with overhead from tensor  transformations and contraction path translation into SQL join operations in \cite{einsteinSQL2023}.}
    \label{fig:time_qcreate}\Description[]{}
\end{figure*}
% \begin{figure}[t]
%     \centering
%     \begin{subfigure}[b]{0.46\linewidth}
%         \centering
%         \includegraphics[width=\linewidth]{figures/experiments/0912/QueryCreation_time_W.png}
%         \caption{W state (\textbf{sparse} circuit)}
%         \label{fig:timequery_wstate}
%      \end{subfigure}
%      \hfill
%       \begin{subfigure}[b]{0.46\linewidth}
%         \centering
%         \includegraphics[width=\linewidth]{figures/experiments/0912/QueryCreation_time_GHZ.png}
%         \caption{GHZ (\textbf{sparse} circuit)}
%         \label{fig:timequery_ghz}
%      \end{subfigure}
%     \caption{Runtime comparison of query creation vs. query execution}%: $a)$  circuit GHZ $b)$ Dense circuit QFT}
%     \label{fig:querytime}
%     % \vspace{-6mm}
% \end{figure}
%================================================================%

An interesting observation from Fig.~\ref{fig:wstate_mem} and \ref{fig:ghzstate_mem} is that  RDBMS solutions perform comparably to the SotA baseline using MPS representation for sparse circuits. 
In particular, the in-memory database SQLite demonstrates a small advantage over the MPS baseline. Fig. \ref{fig:ghzstate_mem} shows that this advantage of RDBMS solutions is even more evident in the GHZ state, which is sparser than the W state (cf. Fig.~\ref{fig:sparsity_GW}). 
Notably, as an early-stage study, we utilized the default configurations for all three RDBMS systems, leaving significant room for optimization to further enhance their performance.

In contrast, in Fig.~\ref{fig:mem_qft}, the memory advantage of the RDBMS solutions is not observed for the Quantum Fourier Transform circuit. 
% The y-axis is in logarithmic scale, as memory usage and runtime often scale exponentially with the number of qubits. 
The memory usage of the RDBMS solution increases at a rate comparable to order-n tensor baseline, sometimes even exceeding it. 
This aligns with our space complexity analysis in Table~\ref{tab:representation}, where the relational representation scales as $\mathcal{O}(n \cdot 2^n)$, while the order-n tensor representation scales as $\mathcal{O}(2^n)$. 
The primary reason is that the QFT circuit mainly involves dense tensor computation.  As shown in Fig.~\ref{fig:sparsity_GW}, the final state of QFT is significantly less sparse compared to the W and GHZ states. 
Besides the final state, the intermediate states generated during QFT circuit simulations are also dense. 
% , requiring large intermediate tensors during tensor contraction. 
% , and the overhead of database storage and access negates any potential benefits. 
% , indicating that the memory advantage of the RDBMS solutions may not be ineffective for dense computations.

% \rihan{@Littau can you please add  2-3 sentences about Fig.~\ref{fig:sparsity_GW} in the right place?}

We further explore the scalability advantages of RDBMS solutions compared to the general baseline using order-n tensor representation.
Fig.~\ref{fig:memlimit} compares the maximum number of qubits that can be simulated by each method under a 2.0 GB memory limit.
% To further investigate the impact of memory constraints, experiments with a $1.0$GB memory limit were conducted. 
For the W state preparation circuit in Fig.~\ref{fig:memlimit_wstate}, 
the RDBMS solutions (SQLite, PostgreSQL, DuckDB) significantly outperform the order-n tensor baseline, in the best case (DuckDB), supporting over $53017$ qubits compared to just 17 qubits for the baseline. This represents a $3118\times$ improvement in scalability. 
As shown in Fig.~\ref{fig:sparsity_GW}, GHZ state preparation circuit, in general, involves even sparser tensors than W state preparation circuit. 
For GHZ state preparation circuit, SQLite performs the best and simulates up to $1.4\cdot10^7$ qubits, compared to 17 for order-n tensor baseline, yielding an $8.3\cdot10^5 \times$ improvement.\footnote{The exact qubit counts are 14107345, 3079324, and 4000862 for SQLite, PostgreSQL, DuckDB, respectively.} 
These results demonstrate the substantial memory efficiency and scalability advantages of RDBMS solutions in handling sparse circuits. %when the number of qubits can be simuluated scales up.
For the dense circuit QFT, RDBMS solutions simulate up to 14 qubits, which is comparable to the 16 qubits supported by order-n tensor baseline with a reduction of $14\%$. 
% This reflects the inefficiency of RDBMS solutions in handling dense circuits. 

% given $2.0$GB memory limit, order-n tensor baseline could simulate up to 17 qubits. while PostgreSQL, DuckDB and SQLite implementations could simulate up to $16\cdot10^{3}$ qubits, representing an increase of almost a thousand times in the number of simulated qubits under constrained memory. The GHZ state preparation circuit which is even sparser showed almost six magnitudes larger in the amount of simulated qubits when comparing the order-n tensor NumPy baseline. This showcases the significant memory-efficiency advantage of database systems in sparse scenarios, when the number of qubits scale up. \\
% For dense QFT circuits, NumPy baseline achieved 16 qubits within the $2.0$ GB memory limit, whereas database implementations could simulate only 14 qubits -- a reduction of 14\%. 
% This reduction underscores the limitations of database systems in handling dense tensor operations under memory constraints.\\ Finally, as a comparison of the naive NumPy implementation to the MPS method it is clear MPS is handling both dense and sparse circuits within reasonable memory usages as shown in \ref{fig:npvsmps}. However, MPS is typically approximated and breaks down when entanglement is high, whereas both naive implementations are fully accurate.

\subsubsection{Runtime}\label{sssec:time}
Figure \ref{fig:wstate_time} compares the runtime of the RDBMS solutions with the baselines. The observations are, in general, similar to memory usage. RDBMS solutions outperform the order-n tensor baseline for sparse circuits (W and GHZ state preparation) but not for the dense circuit QFT. This is consistent with the time complexity analysis in  Table~\ref{tab:time}. 
% The execution time results reveal a nuanced tradeoff between sparse and dense circuits. 
Notably, runtime is influenced by many factors including implementation details. 
For instance, for W state and GHZ state preparation circuits, the SQLite solution shows runtime advantages over both baselines, while the benefits of PostgreSQL and DuckDB over the order-n tensor baseline, become evident only for larger qubit counts (above 23 qubits). These results suggest that to further enhance performance, the optimizations might also need to tailored to individual database systems. 
As shown in figure \ref{fig:qft_time}, the QFT circuit, which involves dense tensor computation, exhibits a different trend. Beyond 12–14 qubits, all RDBMS solutions become slower than both baseline, which is as expected from Table~\ref{tab:time}. Moreover, the baselines are implemented in NumPy, which handles dense tensor operations more efficiently than the RDBMS solutions. This indicates the need for further improvements in both representation and implementation to better simulate dense circuits using RDBMSs.
% compared to specialized numerical libraries. However, this also shows a potential for improvement on the database side, since at least some of the reasons for bad performance lie in the naive data representation used in this implementation. While both implementations show an exponential increase over the number of qubits, the increase for RDBMS implementations is even higher as previously shown in the theoretical time complexities in table \ref{tab:time}.

% \rihan{@Littau could you please change all y-axis captions sentence case? For instance, Memory, Time, ...}

\para{Summary} For Q1, pushing simulation workloads to RDBMS appears to be a promising approach.  For circuits with sparse tensors (W and GHZ), RDBMS solutions demonstrate significant memory and time efficiency, enabling simulations of much larger qubit counts compared to the order-n tensor baseline, while being competitive with the state-of-the-art MPS baseline. 
Further optimization opportunities remain for even better performance. For circuit with dense tensors (QFT), advanced relational representations (beyond Sec.~\ref{sec:rel}), improved system implementations and more sophisticated optimizations are necessary to address current limitations.

\mybox{\textbf{Takeaways:}  
\begin{enumerate}[left=0pt]
    \item\textbf{T1 RDBMS for sparse circuits:} \\
Relational representations are efficient and scalable for quantum circuits involving sparse tensor computations. They are competitive with, and may even outperform, state-of-the-art representations like MPS in specific scenarios.
     
    \item \textbf{T2 Opportunities for Optimization:} \\
    Exciting research opportunities remain, to optimize existing RDBMSs for simulation workload.
\end{enumerate}
}

\subsubsection{Input query generation}
\label{ssec:q_gen}
The input to a simulator is the quantum circuit to be simulated. For RDBMS-based simulators, input circuits can be represented as SQL queries. 
Our preliminary experiments identified the following research gap in the state-of-the-art  approach \cite{einsteinSQL2023} for generating the input circuit SQL queries: 
% \floris{suddenly, [21] seems to be the approach we are investigating? What is actually the difference between our approach and [21]. This may be better clarified early on, in terms of overlap/differences.... }
% \rihan{done, see B1, footnote 11, also in the letter R3D1}

\vspace{0.2cm}
\begin{myframe}[roundcorner=3pt,innerleftmargin = 5pt]
{\bf Q1.} \textit{Should we push the simulation workload to existing DBMSs?  
\begin{enumerate}[label=(\alph*)]
\setcounter{enumi}{1}
 \item How good is the state-of-the-art SQL query generation method for quantum circuit simulation?
\end{enumerate}
 }
\end{myframe}
\vspace{0.2cm}
% \vspace{0.2cm}
% \begin{myframe}[roundcorner=3pt,innerleftmargin = 5pt]
% {\bf 2.} \textit{How good are the state-of-the-art SQL query generation methods for quantum circuit simulation?} 
% \end{myframe}

To generate the SQL queries for circuit simulation, the SotA approach \cite{einsteinSQL2023} transforms tensors representing qubit states and gates— initially represented as NumPy arrays—into a relational representation using a Coordinate (COO) tensor format\footnote{\url{https://docs.scipy.org/doc/scipy/reference/generated/scipy.sparse.coo_matrix.html}}.   
% \floris{Is that representation different from the one we propose? If not, we should make that clear. If yes, also mention.}
Furthermore, the tensor contraction paths also need to be translated into a sequence of join operations in SQL.

Fig. \ref{fig:mem_qcreate} compares the memory usage of  SQL query generation using the 
approach in \cite{einsteinSQL2023} against running the generated query. 
Across all cases, query generation incurs significant memory overhead.
The contrast to query execution is more obvious for sparse circuits like W and GHZ state preparation, where the RDBMS solutions are more efficient. 
The overhead of query generation primarily arises from transforming tensors. 
%For each quantum gate, the corresponding tensor representing it remains the same across different circuits.  
Besides the number of qubits, the memory cost of tensor translation mainly depends on the number and variety of tensors representing gates.  For instance, while the W and GHZ  state circuits 
(Fig~\ref{fig:ghz-w-qft}) share the same number of CONT gates, the W state circuit also includes a number of $G(p)$ gates, which require additional memory for their tensor translation, as seen in Fig.~\ref{fig:mem_qcreate}a and b.
Moreover, the tensor contraction paths need to be translated into a sequence of JOIN operations in SQL.  As the number of qubits increases, the gate count and circuit depth grow, further amplifying the overhead of contraction path translation.

% This translation effectively converts the tensor network or quantum circuit into its relational representation, which is necessary for simulation within an RDBMS framework.

% The memory cost of tensor translation depends on the variety of tensors involved. For circuits like the GHZ state preparation, this cost grows slowly with the increasing number of qubit, as the tensor representation does not change with an increasing number of qubits. In contrast, for the W state preparation circuit, the memory usage scales linearly with the number of qubits, as each added qubit introduces a new variant of the $G(p)$ gate, which must also be translated.

% As shown in figure \ref{fig:qft_time}, dense ciruits, such as the QFT circuit, exhibited a different trend. Numpy's tensor contraction implementation outperformed PostgreSQL, DuckDB and SQLite, primarily due to its efficient handling of dense tensor operations without additional overhead and the COO representation that is used in RDBMS. This highlights that while database systems offer scalability for sparse scenarios, their performance deteriorates for dense circuits compared to specialized numerical libraries. However, this also shows a potential for improvement on the database side, since at least some of the reasons for bad performance lie in the naive data representation used in this implementation. While both implementations show an exponential increase over the number of qubits, the increase for RDBMS implementations is even higher as previously shown in the theoretical time complexities in table \ref{tab:time}.

Fig.~\ref{fig:time_qcreate} compares the query generation time using \cite{einsteinSQL2023} with query execution time. SQL query generation involves constructing a string representation of the query, with the string length increasing proportionally to the number of gates in the quantum circuit, leading to higher time costs.  
% The time cost grows the number of gates.
%
The time overhead is also attributable to two aforementioned factors: the translation of tensors into COO-formatted database relations and the conversion of the contraction path into a sequence of join operations. Each of these two steps introduces computational effort that grows 
%linearly 
significantly with the number of gates and circuit depth. 
%However, this overhead is separate from the actual contraction operation, which represents the core simulation task.

% \mybox{\textbf{Takeaways:}  
% \begin{enumerate*}
%     \item \textbf{T3 SOTA SQL query generation methods for quantum circuit simulation is inefficient.} \\
    
% \end{enumerate*}
% }

\para{Summary} The state-of-the-art SQL query generation method  \cite{einsteinSQL2023} for quantum circuit simulation is inefficient, especially or sparse circuits like W and GHZ states. Possible improvement strategies include pre-storing the tensors in the database, eliminating the need for runtime translation and enabling direct access for quantum circuit operations.  Future improvements, such as optimizing the contraction path generation process, could also reduce the query generation overhead.

% Given that the memory overhead for query creation is distinct from the actual tensor contraction operations, we evaluate the memory usage and performance of the contraction phase independently of the query generation phase.

% \input{Littau}
\begin{table*}[ht!]
\centering
    \captionof{table}{Comparison of simulation with and without databases technology}
    \label{tab:comparison}
    % \vspace{-0.3cm}
            \begin{tabular}{|p{3.5cm}|p{6cm}|p{6cm}|}
                \hline
                \rowcolor[gray]{0.9} % Light gray background for the header row
                \textbf{Feature} & \textbf{Without Databases} & \textbf{With Databases} \\
                \hline
                \textbf{Data Storage \& Retrieval} & 
                Data, such as simulation configurations, intermediate and final results, is stored in flat files (e.g., HDF5, JSON). Retrieval requires custom scripts and manual searches, which hinders data discovery, mining, and analysis. & 
                Data stored in structured, %tensor based, 
                indexable formats, enabling efficient organization and querying; optimized retrieval through indexing, caching, and execution plans, 
                %greatly enhancing data access performance 
                and facilitating the use of existing tools for data discovery, mining, and analysis.\\
             \hline
                \textbf{Query Languages} & 
                Accessing data relies on low-level programming, inefficient for querying data. & 
                High-level query languages like SQL, relational algebra, or tensor languages simplify data retrieval by providing intuitive, abstract constructs. These languages allow users to perform efficient and complex queries without requiring low-level programming, optimizing the interaction with large datasets.
                % High-level query languages (e.g., SQL, relational algebra, or Tensor Languages) simplify retrieval, enabling efficient queries.
                % for tensor structures, gate sequences, and circuit outputs. 
                \\
                \hline
                \textbf{Consistency, Recovery, and Transactions} & 
                Manual effort or customized scripts, error-prone. & 
                Well-developed theory and tools, reducing human effort and possible errors. \\
                \hline
                \textbf{Scalability} & 
                Limited by hardware capacity. & Potential benefits of databases for memory-efficient simulations; 
                extended by using secondary storage and out-of-core database technologies\\
                \hline
                % \textbf{Integration} & 
                % Custom solutions needed with manual overhead, limited commercial tool suppor. & 
                % Easy integration with existing DBMSs and their ecosystems. \\
                % \hline
            \end{tabular}
\end{table*}

% \newpage
\section{More data management opportunities}
In this section, we continue with Sec. \ref{ssec:opp} and list more research opportunities. 

%adjust footnote
% \enlargethispage{\baselineskip}
 \para{NoSQL databases \& Quantum data management}
NoSQL databases offer a promising solution for analyzing simulations and quantum experiments, since NoSQL databases are well-suited for managing unstructured data with their scalability and flexible schemas. 
% They include key-value stores, document stores, column-family databases, and graph databases, with the capability of horizontal scaling, making them ideal for large datasets. Their flexible schemas adapt easily as models evolve, which is crucial for dynamic simulations. 
A forward-thinking extension of Q1 in Sec.~\ref{ssec:dbq2} is to choose most suitable databases for simulation. For example, TileDB \cite{TileDB}, an array database, offers native storage and retrieval of multi-dimensional arrays (tensors), leveraging parallel processing, distributed computing, and easy integration with machine learning and data science tools. Adapting array databases to simulation workloads remains an open challenge, requiring significant effort to enable in-database computation and query optimization for simulation workload and expand the their ecosystem to match the versatility of RDBMSs.

% TileDB  supports HDF5 files, enabling the conversion of HDF5 datasets into TileDB arrays for more efficient queries and storage.
Moreover, JSON is a widely used file type for circuit descriptions and simulation configurations in quantum frameworks, e.g., Qiskit \cite{Qiskit}, Cirq \cite{cirq}, or Braket \cite{Braket}.  
Document stores like MongoDB \cite{MongoDB} can efficiently manage JSON and could play a critical role in this context. 
% For instance, key-value stores such as Redis and DynamoDB are optimized for low-latency access for simulation workloads requiring low latency.   These features make NoSQL a strong choice for large-scale, low-latency, flexible simulation data management. 
Furthermore, for error correction techniques discussed in Sec. 4.3, exploring the recent effort of graph analytics \cite{10.14778/3611540.3611577, yan2005substructure, 10.1145/3448016.3452826, 
10.1145/2627692.2627694,
behnezhad2023exponentially} and graph databases \cite{tian2023world} such as Neo4j\footnote{\url{https://neo4j.com/}},  TigerGraph\footnote{\url{https://www.tigergraph.com/}}, might yield valuable insights.  

In summary, relational and NoSQL databases, each offer unique advantages, but no single database system is likely to meet the diverse requirements of all quantum applications; the optimal choice will depend on the specific needs of each quantum application.

\section{More discussion on simulation with database technologies}

Continuing with Sec. 4, we compare simulation without and with databases in Table~\ref{tab:comparison}.
% We conducted extensive discussions with quantum theorists and simulation engineers to understand their practical needs and identified potentially useful database functions in Table 1. A key takeaway is that quantum simulation applications vary significantly regarding data sizes, computation complexity, and user demands. These variations suggest that different database technologies may be more effective for specific scenarios (exact or approximate simulations, applications such as quantum supremacy or quantum networks, etc.). Currently, database technologies remain underutilized in the simulation landscape, highlighting substantial opportunities for future research.

\emph{1. Data Storage \& Retrieval.} A common need in many simulation applications is managing data effectively, particularly optimizing retrieval and supporting advanced data analytics. Existing simulation tools often rely on flat files to store configurations and results, requiring custom scripts (e.g., Python) or manual searches for access. This approach is inefficient, time-consuming, and limits data discovery, mining, and analysis. For instance, in quantum network simulations, distributed experiments generate data from multiple users, but lack efficient tools for data discovery and analytics.
In contrast, databases store data in structured, indexable formats, enabling efficient organization and fast querying. Optimized retrieval through indexing, caching, and execution plans enhances performance. Another benefit is that database technologies are widely used and have a rich ecosystem of commercial and open-source tools (e.g., data analytics, discovery, integration, and cleaning platforms) that support advanced analysis and integration with existing simulation workflows.

\emph{2. Query Languages.} Current simulation tools typically access data relying on programming languages (e.g., Python) requiring customized optimizations for different quantum applications. This approach significantly increases development overhead and limits flexibility. In contrast, 
database systems leverage high-level query languages like SQL, relational algebra, or tensor-specific languages \cite{DBLP:journals/pvldb/KoutsoukosNKSAI21, geerts2021matrix, brijder2019matlang} to simplify data retrieval. These languages allow users to perform complex queries in a declarative and efficient manner, facilitating more effective querying and more sophisticated analyses with less human effort.

\emph{3. Consistency, Recovery, and Transactions.} Please refer to our response in D2/R2.

\emph{4. Scalability.} We have discussed the bottleneck of scalability without databases in Sec.  4.1. With Fig.~\ref{fig:mem} and \ref{fig:memlimit}, we identified the potential benefits of database-based approaches for memory-efficient simulations. Another promising direction is out-of-core simulation using databases \cite{Trummer24}. 
When data sizes exceed memory limits, most main-memory databases switch to out-of-core strategies to complete queries. However, this often results in performance that is significantly slower than in-memory processing. %orders of magnitude slower
Recent advancements in out-of-core database techniques, such as sorting \cite{10184754}, aggregation \cite{10597735}, and joins, have redesigned operators to ensure performance degrades significantly less when the data size exceeds memory limits. It is an intriguing research direction to explore whether these techniques can enhance out-of-core simulation when simulation workloads are integrated with database systems.

\end{appendices}

% %\onecolumn
% % \appendix
% % \input{appendix.tex}

\end{document}